\documentclass[aps,pra,showpacs,twocolumn,superscriptaddress]{revtex4-1}
\usepackage{amssymb,latexsym,mathrsfs}
\usepackage{amsfonts}
\usepackage{graphicx} 
\usepackage{lmodern} 
\usepackage{fix-cm}
\usepackage{xfrac}
\usepackage{color}    
\usepackage{textcomp}
\usepackage{amsmath}
\usepackage{float}
\usepackage{datetime}
\usepackage{cancel}
\usepackage{hyperref}
\hypersetup{
    linkcolor=red,
    filecolor=magenta,      
    urlcolor=cyan,
}
\usepackage{tabularx,ragged2e,booktabs}

\newcommand{\bearr}{\begin{array}}
\newcommand{\enarr}{\end{array}}

\newcommand{\bea}{\begin{eqnarray}}
\newcommand{\eea}{\end{eqnarray}}
\def \be{\begin{eqnarray}}
\def \ee{\end{eqnarray}}

\begin{document}

\title{Generating Entanglement by Quantum Resetting}

\author{Manas Kulkarni}
\email{manas.kulkarni@icts.res.in}
\affiliation{International Centre for Theoretical Sciences, Tata Institute of Fundamental Research, Bengaluru -- 560089, India}

\author{Satya N. Majumdar}
\email{satya.majumdar@universite-paris-saclay.fr}
\affiliation{LPTMS, CNRS, Univ.  Paris-Sud,  Universit\'e Paris-Saclay,  91405 Orsay,  France}

\date{\today}

\begin{abstract}

We consider a closed quantum system subjected to stochastic Poissonian resetting with rate $r$ to its initial state.
Resetting drives the system to a nonequilibrium stationary state (NESS) with a mixed density matrix which
has both classical and quantum correlations.
We provide a general framework to study these NESS correlations for a closed quantum system with a general
Hamiltonian $H$. We then apply this framework to 
a simple model of a pair of ferromagnetically coupled spins, starting from state $\mid\downarrow\downarrow \rangle$ and resetting to the same state with rate $r$.
We compute exactly the NESS density matrix of the full system. This then provides access to three basic observables, namely (i) the von Neumann entropy of a subsystem (ii) the fidelity
between the NESS and the initial density matrix and (iii) the concurrence in the NESS (that provides a measure
of the quantum entanglement in a mixed state), as a function of the two parameters: the resetting rate and
the interaction strength.
One of our main conclusions is that a nonzero resetting rate and a nonzero  interaction strength 
generates quantum entanglement in the NESS (quantified by a nonzero concurrence) and moreover this concurrence
can be maximized by appropriately choosing the two parameters. Our results show that quantum resetting provides a simple and effective mechanism to enhance entanglement between two parts of an interacting quantum system.

\end{abstract}

\setcounter{page}{1}
\maketitle


\section{Introduction}
\label{sec:intro}

Entanglement is a fundamental property of a quantum system which has no classical counterpart~\cite{NC_Book_2010, BZ_Book_17, AFOV08}. A pure state of a bipartite quantum system is called entangled if is not factorizable into the states of the subsystems. The simplest example of such an entangled state is a singlet state of a pair of spin $1/2$ particles: $|\psi\rangle =(\mid \uparrow \downarrow \rangle - \mid \downarrow \uparrow \rangle )/\sqrt{2}$. For a pure state, a simple measure of the degree of entanglement, i.e., the lack of factorization is provided by the von Neumann entanglement entropy~\cite{NC_Book_2010, BZ_Book_17, AFOV08} of a subsystem with the rest. This entropy has been studied extensively for many quantum systems with applications ranging from quantum information theory and quantum cryptography to many body quantum condensed matter systems~\cite{P93,W98,CC04,CC05,AFOV08,FMPPS08,NMV10, SNM10,NMV11,SLRD13,CDM15,KTLRSPG16,BEJVMLZBR19,LMS19,CP20,SKCD21,FG21,CTKC22}.  Recently, there have been enormous interest in studying the entanglement in quantum systems driven out of equilibrium, either by a sudden quench of a parameter of the Hamiltonian~\cite{CC05,EP07,C11,AC18,A18}  or by repeated projective measurements \cite{CTKC22, LCF19, SRN19,GH20,ZGWGHP20,RCGG20}. Furthermore, developing and designing protocols that enhances quantum entanglement is central to quantum information theory and has been of enormous interest in recent years~\cite{NPSC12, AKH16,SMFAKTS16,AKT14,WWW19,RSMMVEKWSS14,SSHCHMMGA17,STT16,DBVDCK13}. 

If however the quantum system is in a mixed state, it is not easy to distinguish the quantum correlations from classical correlations between two subsystems. In this case, the von Neumann entropy is not well suited to characterise the `purely quantum nature' of the correlations between two subsystems since it contains information about quantum correlations as well as of classical probabilities. Several measures have been proposed in the literature to characterise the `quantumness' of the correlations. This includes concurrence~\cite{W98,HW97, HHHH09}, quantum discord~\cite{OZ01,GG10, GTA13,HV21} etc. Amongst these, concurrence is one of the most widely used measures of entanglement in a mixed state and has been extensively investigated~\cite{W98,HW97,W01,MKB04, HHHH09}. It is worth emphasising that even though
von Neumann entropy of a subsystem is not a suitable measure to compute entanglement (or more generally quantum correlations) for a mixed state, it is however a crucial ingredient to compute some of these other measures of quantum correlations such as the
quantum discord~\cite{OZ01,GG10, GTA13,HV21}. Finally, 
another interesting quantity is fidelity~\cite{NC_Book_2010, BZ_Book_17} which measures the ``distance" between two states (mixed or pure) -- in particular, it can be used to measure the closeness of a mixed state  to a reference pure state. 

Recently, `quantum resetting' has been proposed~\cite{MSM18,RTLG18} as a simple protocol to drive a quantum system out of equilibrium. Under this protocol, the unitary evolution of a quantum system starting from its initial state $|\psi (0) \rangle$ is interrupted at random Poissonian times with rate $r$ and the system is instantaneously reset to its initial state $|\psi  (0)\rangle$. Between two successive resets the system evolves unitarily. It has been shown that this repeated resetting at random times drives the system to a non-equilibrium steady state (NESS) where the density matrix acquires non-zero off-diagonal elements~\cite{MSM18}. 
In addition, the density matrix typically becomes mixed in the NESS, due to the fact that resetting induces `classical' 
probabilities in the density matrix.
Over the last decade, the effect of resetting has been studied extensively in a wide variety of classical systems evolving via stochastic dynamics~\cite{EM11,EM12} (for reviews, see Refs.~\cite{EMS20, PKR22,NG23}). Poissonian resetting to the initial condition in such systems manifestly breaks detailed balance and typically
drives the system to a NESS. The resulting classical NESS has been characterised in a variety of theoretical models~\cite{EM11,EM12, EM_14,GMS_14,MSS_15,CS_15, MV_16,MMS_20,BLMS_23, MV13,CM16,EM16,EM18,MM19,P15,MB18,BCS19}, as well
as in experiments~\cite{TPSRR_20,BBPMC_20, FBPCM_21} involving diffusing colloids in an optical trap. While there have been few recent studies on quantum resetting~\cite{MSM18,RTLG18,PCML21,PCL22,DDG21,MCPL22,DCD23,SV23,YB23.1,YB23.2},
there are only few recent studies on the effect of resetting on correlations in quantum systems ~\cite{MCPL22,TDFS22}.
For example, does resetting
increase or decrease the quantum entanglement between two subsystems? Can one characterise the entanglement between two subsystems in the resetting induced NESS in a quantum system? 

In this paper, we investigate three quantities, namely (i) von Neumann entropy of a subsystem (ii) fidelity and (iii) concurrence
in the resetting induced NESS of an isolated quantum system with a generic Hamiltonian. We then apply this framework
in a very simple model consisting of a pair of ferromagnetically interacting spins (qubits) with coupling strength $J$ in the presence of a transverse magnetic field and subjected to Poissonian resetting 
with rate $r$ to its initial state (which for simplicity is assumed to be a pure state $\mid\downarrow\downarrow\rangle$). We show that all the three quantities (i), (ii) and (iii) can be computed exactly in the resetting
induced NESS where the density matrix is mixed. Our results show rather rich and interesting dependence of these
quantities on the resetting rate $r$ and the coupling $J$ between the spins. One of our main conclusions is that a nonzero resetting rate $r$ and a nonzero  interaction strength $J$
generates quantum entanglement in the NESS (quantified by a nonzero concurrence) and moreover this concurrence
can be optimized by appropriately choosing the two parameters $R$ and $\alpha$.

The rest of the paper is organised as follows. In Section \ref{sec:genf}, we introduce the general framework to study
quantum correlations between two subsystems in an isolated quantum system with a generic Hamiltonian $H$ subject to Poissonian resetting. In Section \ref{sec:tls},
we present a simple model of a ferromagnertically coupled spin pairs and calculate explicitly the von Neumann entropy
of the first spin (Section \ref{vne}), fidelity between the NESS and the initial pure state (Section \ref{fid}) and the concurrence (Section \ref{conc}) that measures the quantum entanglement between the two spins in the NESS.
Finally we conclude in Section \ref{sec:conc}. Some details of the calculations are relegated to the appendix.

\section{General Framework}
\label{sec:genf}

Consider any isolated quantum system with a time-independent Hamiltonian $H$ whose eigenstates are denoted by $|E\rangle$ with associated eigenvalue $E$. The system is prepared initially in a pure state $|\psi(0)\rangle$ (which is not an eigenstate of $H$). Consequently the density matrix is given by $\hat{\rho} (0)=  | \psi(0)\rangle \langle \psi(0)|$ with $\rm{tr}[\hat{\rho} (0)] =1$. In a closed quantum system the state evolves unitarily via the Schrodinger equation $|\psi(t)\rangle = e^{-iHt} | \psi(0)\rangle$. Consequently the density matrix $ \hat{\rho} (t)=  | \psi(t)\rangle \langle \psi(t)|$ evolves via $\hat{\rho} (t) = e^{-iHt} \hat{\rho} (0)e^{iHt}$. This unitary evolution preserves the trace, i.e., $\rm{tr}[\hat{\rho} (t)] =1$ for any $t$. In Ref.~\cite{MSM18}, the protocol of quantum resetting was introduced where the state of the system 
evolves in time by a mixture of deterministic unitary dynamics and stochastic classical resetting moves. More precisely, the system evolves in continuous time $t$ according to the stochastic rule
\begin{equation}
|\psi(t+dt)\rangle =\begin{cases} (1-iH dt)|\psi(t)\rangle\,, \,\,\, \text{with prob.} \,\,\, 1-r\, dt \\ 
\\
|\psi(0)\rangle \,, \quad  \quad \quad \quad \quad  \,   \text{with prob.} \,\,\, r\, dt\, ,\end{cases}
\label{eq:reset}
\end{equation}
where $r$ represents the resetting rate. Under this resetting dynamics, it was shown that the density matrix evolves as~\cite{MSM18}
\begin{equation}
\hat{\rho}_r(t)  = e^{-rt} \hat{\rho}(t) + r\int_0^t d\tau e^{-r \tau}  \hat{\rho}(\tau)\, ,
\label{eq:reset1}
\end{equation}
where the subscript $r$ in $\hat{\rho}_r(t)$ indicates a finite resetting rate and $\hat{\rho} (t) = e^{-iHt} \hat{\rho} (0)e^{iHt}$ is the time evolved density matrix in the absence of resetting. Note that for any finite $t>0$, even though $\hat{\rho}(t)$ (in the absence of resetting) represents a pure density matrix, the resetting induced density matrix $\hat{\rho}_r(t) $ on the left hand side of Eq.~\eqref{eq:reset1} is generically mixed for $r>0$. As time $t\to \infty$,
the density matrix $\hat{\rho}_r(t) $ approaches a non-equilibrium steady state (NESS) given by 
\begin{equation}
\hat{\rho}_r(\infty)  =  r\int_0^\infty d\tau e^{-r \tau}  \hat{\rho}(\tau)\, . 
\label{eq:reset1_ness}
\end{equation}
Thus the NESS density matrix $\hat{\rho}_r(\infty)$ is {\em mixed} and can be viewed as $r$ times the Laplace transform (with respect to $\tau$) of the density matrix $\hat{\rho}(\tau)$ without resetting. Computing the NESS density matrix with resetting in Eq.~\eqref{eq:reset1_ness} thus requires the full knowledge
of the density matrix $\hat{\rho}(\tau)$ of the system without resetting {\em at all times} $\tau$, and this is typically nontrivial.
It turns out that the NESS density matrix with resetting can be written explicitly in the energy basis as~\cite{MSM18}
 \begin{equation}
\hat{\rho}_r(\infty)  =  \begin{cases}   \rho_{E,E}(0),\, \,\,\,\quad \quad \quad \quad \quad \text{if} \,\,  E=E'\\
 \rho_{E,E'}(0) \frac{r}{r+i(E'-E)}, \, \quad \text{if} \,\, E\neq E'
\end{cases}
\label{eq:reset1_ness1}
\end{equation}
In Eq.~\eqref{eq:reset1_ness1}, the subscript ($E,E'$) denotes the elements of the initial density matrix $\hat{\rho} (0)$ in the energy basis. Thus, in the presence of resetting, the density matrix $\hat{\rho}_r(\infty)$ acquires nonzero off-diagonal elements. Note that if one takes the $r\to 0^+$ limit, the off-diagonal elements in Eq.~\eqref{eq:reset1_ness1} vanish and the system approaches a stationary density matrix with only diagonal elements in the energy basis. However, we note that this is not the same if one keeps $t$ finite and takes the $r \to 0 $ limit. In that case, the system does not reach a stationary state as the off-diagonal elements keep oscillating in time. Thus, the two limits $\lim_{t\to \infty} \lim_{r\to 0}$ and $\lim_{r\to 0} \lim_{t\to \infty}$ do not commute. 

Given the exact density matrix at time $t$ in Eq.~\ref{eq:reset1} in the presence of resetting, one can, in principle, compute various observables of interest at any finite time $t$ and in particular, in the steady state. The goal of this paper is to investigate (i) von Neumann entropy of a subsystem in the NESS, (ii) fidelity between the density matrix in the NESS ($t\to \infty$) and the initial density matrix $(t=0)$
and (iii) concurrence in the NESS that quantifies the entanglement between two subsystems in a mixed state. The definitions of these quantities are provided below.

\vskip 0.3cm
\noindent
\textbf{von Neumann entropy:}
In order to compute the von Neumann entropy of a subsystem $A$ of the full system, we need to first compute the reduced density matrix of the subsystem $A$ by tracing out the degrees of freedom belonging to $\bar{A}$ which is the complement of $A$, i.e., 
\begin{equation}
\hat{\rho}_{A,r} (t)= \rm{tr}_{\bar{A}} \big[ \hat{\rho}_{r} (t)\big]\, ,
\label{eq:partial_rho}
\end{equation}
where $\hat{\rho}_{r} (t)$ is given in Eq.~\eqref{eq:reset1}. The von Neumann entropy is then defined as 
\begin{equation}
S_r(t) =  - \rm{tr}\big[ \hat{\rho}_{A,r} (t) \ln(\hat{\rho}_{A,r} (t))\big] = -\sum_{i=1}^{N_A} \lambda_i (t) \ln(\lambda_i (t))\, , 
\label{eq:vnent}
\end{equation}
where $\lambda_i(t)$ are the eigenvalues of $\hat{\rho}_{A,r} (t)$ and $N_A$ is the size of the subsystem $A$. 
As mentioned in the introduction, even though the von Neumann entropy is not a suitable measure of quantum entanglement in
a mixed state [such as in the resetting induced NESS in Eq.~(\ref{eq:reset1_ness})], it is nevertheless useful to compute this
entropy as this is a crucial ingredient to build other measures of quantum correlations such as the quantum discord. 
By performing the partial tracing over $\bar{A}$ directly in Eq.~(\ref{eq:reset1}), one obtains the evolution equation
for the reduced density matrix $\hat{\rho}_{A,r}$, 
\begin{equation}
\hat{\rho}_{A,r}(t)  = e^{-rt} \hat{\rho}_A(t) + r\int_0^t d\tau e^{-r \tau}  \hat{\rho}_A(\tau)\, .
\label{eq:reset1_partial}
\end{equation}
Hence, one sees that as $t\to \infty$, the reduced density matrix also approaches a stationary limit as $t\to \infty$
\begin{equation}
\hat{\rho}_{A,r}(\infty)  =  r\int_0^\infty d\tau e^{-r \tau}  \hat{\rho}_A(\tau)\, .
\label{eq:reset1_partial_ness}
\end{equation}
Consequently, the von Neumann entropy also approaches a stationary limit as $t\to \infty$
\begin{equation}
S_r(\infty) =  - \rm{tr}\big[ \hat{\rho}_{A,r} (\infty) \ln(\hat{\rho}_{A,r} (\infty))\big] \, ,
\label{eq:vnent_ness}
\end{equation}
where $\hat{\rho}_{A,r} (\infty)$ is given in Eq.~\eqref{eq:reset1_partial_ness}. While the full NESS density matrix has a simple
explicit form in the energy basis in Eq.~(\ref{eq:reset1_ness1}), it turns out that 
performing the partial trace in Eq.~\eqref{eq:vnent_ness} in the energy basis is rather hard and it becomes easier if one changes to the local basis involving local degrees of freedom (e.g., the site basis on a lattice). This is shown explicitly in the two spin
model discussed in Section \ref{vne}.

\vskip 0.3cm
\noindent
\textbf{ Fidelity:} The fidelity between two density matrices $\hat{\rho}$ and $\hat{\sigma}$ (pure or mixed) provides a measure of the closeness between them. It is defined as~\cite{NC_Book_2010, BZ_Book_17}
\begin{equation}
\mathcal{F} (\hat{\rho},\hat{\sigma}) =  \bigg(\rm{tr}\left[\sqrt{\sqrt{\hat{\rho}}\, \hat{\sigma} \,\sqrt{\hat{\sigma}}}\right]\bigg)^2\, .
\label{fid_def}
\end{equation}
If one of the matrices, say $\hat{\sigma}$ is pure, i.e., $\hat{\sigma} = |\psi_\sigma\rangle \langle \psi_\sigma|$ then the definition in Eq.~\eqref{fid_def} reduces to a simpler expression
\begin{equation}
\mathcal{F} (\hat{\rho},\hat{\sigma}) =  \langle \psi_\sigma| \hat{\rho}  |\psi_\sigma\rangle \, .
\label{fid_def_simp}
\end{equation}
In our case, we assume that the system starts from a pure state $  |\psi (0)\rangle$. Hence the initial density matrix corresponds to a pure state ${\hat \rho}_r(0)= \hat{\rho}(0) =   |\psi (0)\rangle \langle \psi (0) |$. 
It is then natural to ask how close is the NESS to the initial state. This is measured by the fidelity between the density matrix in the NESS 
$\hat{\rho}_r(\infty) $ in Eq.~\eqref{eq:reset1_ness} and the initial density matrix $\hat{\rho}(0)$. Hence, we can use the simplified expression in Eq.~(\ref{fid_def_simp}) by identifying $\hat \sigma= {\hat \rho}_r(0)$ (since this is pure) and $\hat \rho= {\hat \rho}_r (\infty)$ leading to
\begin{equation}
\mathcal{F}(\hat{\rho}_r(\infty), \hat{\rho}_r(0)) =  \langle \psi(0) | \hat{\rho}_r(\infty) | \psi(0) \rangle\, .
\label{fid_def_simp1}
\end{equation}
Eq.~\eqref{fid_def_simp1} holds for a quantum system with arbitrary Hamiltonian $H$.  We will compute this explicitly in the two spin
model discussed in Section \ref{fid}.

\vskip 0.3cm
\noindent 
\textbf{Concurrence:} Concurrence is a well known measure to characterise the quantum entanglement between two subsystems in a mixed state~\cite{HW97,W98,W01,HHHH09}. However, it is extremely hard  to compute this quantity for a closed quantum system with a generic Hamiltonian $H$, since it involves a complex optimization problem in high dimensions~\cite{W98,HW97, MKB04, HHHH09}.
However,  
for a pair of qubits with a mixed density matrix $\hat{\rho}$, there is an explicit expression for the concurrence~\cite{W98,HW97, HHHH09}
\begin{equation}
\mathcal{C}(\hat{\rho}) = \rm{max} (0, \mu_1 - \mu_2 - \mu_3 - \mu_4)\, ,
\label{eq:C1}
\end{equation}
where $\mu_i$'s are the eigenvalues in decreasing order of the matrix  
\begin{equation}
\mathcal{R} =  \sqrt{\sqrt{\hat{\rho}}\, \tilde{\rho} \, \sqrt{\hat{\rho}}} \, ,
\label{eq:C2}
\end{equation}
with $\tilde{\rho}$ defined by 
\begin{equation}
\tilde{\rho} = (\sigma_y \otimes \sigma_y) \, \hat{\rho}^* \,  (\sigma_y \otimes \sigma_y) \, .
\label{eq:C3}
\end{equation}
Here, $\sigma_y = \left(
\begin{array}{cc}
 0 & -i \\
 i & 0 \\
\end{array}
\right)$ is the $y$ component of the Pauli spin matrix and $\hat{\rho}^*$ is the complex conjugate of $\hat{\rho}$. The concurrence $C$ can take values in $C\in [0,1]$. It achieves its maximal value for a fully entangled state. In contrast, it vanishes for any
mixed state which can be expressed as a convex combination of product states, i.e., when the density matrix can be expressed as a convex roof of separable density matrices~\cite{MKB04}
\begin{equation}
\hat{\rho} = \sum_{i} p_i \, \hat{\rho}^A_i \otimes \hat{\rho}^B_i \,\,\, {\rm where}\,\, \,0\le p_i\le 1\,\,\,{\rm and}\,\,\,  \sum_i p_i =1\, . 
\label{eq:sep}
\end{equation} 
Such a mixed state contains classical correlations, but no quantum entanglement as demonstrated by the vanishing of the
concurrence.  For a pure state $\hat{\rho} = |\psi\rangle\langle\psi|$, the formula for the concurrence in Eq. (\ref{eq:C1}) further reduces to~\cite{W98} 
\begin{equation}
\mathcal{C}(\hat{\rho}) = \sqrt{2\,(1-\rm{tr} \hat{\rho}_{A}^2\,)} \, ,
\label{eq:C1_simp}
\end{equation}
where $\hat{\rho}_{A}$ is the reduced density matrix given by $\hat{\rho}_{A} = \rm{tr}_{\bar{A}} \big[ \hat{\rho} \big]$ and $\bar{A}$  is the complement of $A$. In Section \ref{conc}, we will compute the concurrence for the two spin model subject to resetting with rate $r$ where the density matrix $\hat{\rho}$ in Eq.~\eqref{eq:C1} is replaced by the NESS density matrix $\hat{\rho}_r(\infty)$ in Eq.~\eqref{eq:reset1_ness}. 

\section{Two spin model}
\label{sec:tls}
 
We consider a pair of spins with the Hamiltonian~\cite{P70, S_Book_2011, S73,SRULSS23}
\begin{equation}
H= - J \sigma_1^ z  \sigma_2^ z + \frac{\Omega}{2} (\sigma_1^x  + \sigma_2^x) \,,
\label{eq:ham}
\end{equation}
where $\sigma$'s in Eq.~\eqref{eq:ham} are the Pauli matrices, $J>0$ is the ferromagnetic coupling between the spins and $\Omega>0$ is the transverse magnetic field associated with each spin. This is simply the transverse field Ising model with two spins.
Here, our local basis is the Hilbert space composed of the eigenstates of $\sigma_i^z$ where $i=1,2$ label the two spins. This Hilbert space has a dimension $4$ consisting of the basis vectors $\mid\uparrow \uparrow \rangle,\mid \uparrow \downarrow \rangle,\mid \downarrow \uparrow \rangle,\mid\downarrow \downarrow \rangle$. In this basis,  the Hamiltonian is represented by a $4\times 4$ matrix 
\begin{equation}
H= \begin{pmatrix} - J & \Omega/2 & \Omega/2  & 0 \\
      \Omega/2  & J & 0 & \Omega/2 \\
       \Omega/2 &0 & J & \Omega/2  \\
       0 &\Omega/2 &\Omega/2  & -J. 
\end{pmatrix}
\label{eq:ham_mat}
\end{equation}
We prepare the system initially in the state $|\psi(0)\rangle=\mid\downarrow\downarrow\rangle$ and also reset it to $|\psi(0)\rangle$ with rate $r$. We choose this initial and the reset state to be a pure state $\mid\downarrow\downarrow\rangle$ for simplicity, but
our framework can be easily extended to the case when the initial and the rest state is a mixed state, such as a singlet
$ (\mid\uparrow\downarrow\rangle-\mid\downarrow\uparrow\rangle)/\sqrt{2}  $ or a Bell state $(\mid\uparrow\uparrow\rangle+\mid\downarrow\downarrow\rangle)/\sqrt{2}  $. With our choice of the initial state,
the initial density matrix is thus 
\begin{equation}
\hat{\rho}(0)=|\psi(0)\rangle\langle \psi(0)|=\mid\downarrow \downarrow \rangle \langle\downarrow \downarrow \mid \, . 
\label{eq:init_rho}
\end{equation}
In the local basis, this initial density matrix is then represented by a $4\times 4 $ matrix,  $\hat{\rho}(0)= [\{ 0,0,0,0 \}, \{0, 0, 0, 0 \}, \{ 0, 0, 0, 0 \}, \{0, 0, 0, 1 \}] $. We want to calculate the quantum entanglement of spin $1$ with that of spin $2$. Thus, in this case, the spin $1$ represents the subsystem $A$, while the spin $2$ represents its complement $\bar{A}$. 

We note that very recently Magoni et al studied~\cite{MCPL22} a system of $N$ non-interacting spins 
with a Hamiltonian  
\begin{equation}
H=  \Omega \sum_{i=1}^{N} \sigma_i^x + \Delta \sum_{i=1}^{N} \sigma_i^x\,, ,
\label{eq:ham_m}
\end{equation}
starting from the initially all up state and resetting to this state with rate $r$. Interestingly, they showed that even though the spins are
noninteracting, the simultaneous resetting of all spins together induces a correlation between the spins in the NESS.
A similar mechanism for generating strong correlations via simultaneous resetting in a classical system was
demonstrated in Ref.~\cite{BLMS_23} where the authors studied a system of $N$
 independent Brownian motions on a line, starting and resetting simultaneously with rate $r$ to the same position. 

However, the nature of the correlations generated by
simultaneous resetting in a quantum system with a {\em noninteracting} Hamiltonian such as in Eq.~(\ref{eq:ham_m}) still remains `classical' and does
not generate quantum entanglement between the spins. This is seen from the fact that the density matrix
at any time $t$ for the noninteracting Hamiltonian can be expresssed as a convex linear combination of separable density matrices
as in Eq.~(\ref{eq:sep}) 
leading to a vanishing concurrence at all time $t$, including in the NESS. Our main motivation in
this paper is to investigate the effect of resetting on the quantum entanglement, and for that it is crucial
to have an interaction term in the Hamiltonian as in Eq.~(\ref{eq:ham}). Indeed, we will show that in our model,
the concurrence has a nonzero value in the NESS and moreover, it gets maximal in certain regions of the parameter
space. We have three parameters $(J, \Omega, r)$. However, one can express all physical quantities in
terms of only two dimensionless parameters
\begin{equation}
R= \frac{r}{\Omega}\, \quad \alpha=\frac{J}{\Omega}\, .
\label{rescaled_para}
\end{equation}
We will see that the concurrence in the NESS has nontrivial behaviour in the $(R,\alpha)$ plane.
One of our main conclusions is thus: a nonzero resetting rate $R$ and a nonzero interaction strength $\alpha$ are both
crucial to generate and enhance quantum entanglement in the NESS.

\subsection{ von Neumann entropy:}
\label{vne}

In the presence of resetting, the reduced density matrix of a spin, say the spin $1$ at time $t$ is given by Eq.~\eqref{eq:reset1_partial}. 
Thus, we need to first evaluate the reduced density matrix $\hat{\rho}_A(t)$ of the subsystem $A$, i.e., the spin 1 without resetting. This can be evaluated as follows. Without resetting the full density matrix evolves by $\hat{\rho} (t) = e^{-iHt} \hat{\rho} (0)e^{iHt}$. 
In the local basis this can be represented as $ \hat{\rho} (t)  =  \sum_{i\alpha j\beta} \rho_{i\alpha,j\beta} (t) |i\alpha\rangle \langle j\beta|$ where the index $i,j$ refers to the states of spin 1, while  $\alpha,\beta$ labels the states of spin 2. Hence, $\hat{\rho}_A(t) =  \rm{tr}_{\bar{A}} \big[ \hat{\rho}_{r} (t) \big]= \sum_\alpha \langle \alpha | \hat{\rho}(t) | \alpha \rangle$. Using the matrix representation of $\hat{\rho} (t)$ one then gets $\hat{\rho}_A(t) = \sum_{ij}  \Big[ \sum_\alpha \rho_{i\alpha,j\alpha} (t)  \Big] |i\rangle \langle j|$. Thus $\hat{\rho}_A(t)$ is a $2\times 2$ matrix given by
\begin{equation}
\hat{\rho}_A(t) = \begin{pmatrix}
\rho_{\uparrow\uparrow,\uparrow\uparrow}(t) + \rho_{\uparrow\downarrow,\uparrow\downarrow}(t) & \rho_{\uparrow\uparrow,\downarrow\uparrow}(t) + \rho_{\uparrow\downarrow,\downarrow\downarrow}(t) \\
\rho_{\downarrow\uparrow,\uparrow\uparrow}(t) + \rho_{\downarrow\downarrow,\uparrow\downarrow}(t) & 
\rho_{\downarrow\uparrow,\downarrow\uparrow}(t) + \rho_{\downarrow\downarrow,\downarrow\downarrow}(t) 
\end{pmatrix}\, .
\label{eq:rhoAt}
\end{equation}
Using the matrix representation of $\hat{\rho} (0)$ and $H$, one can evaluate the matrix elements of  $\hat{\rho} (t) = e^{-iHt} \hat{\rho} (0)e^{iHt}$ using the Mathematica. Consequently, the elements of the reduced density matrix in Eq.~\eqref{eq:rhoAt} can be obtained explicitly.  In terms of the two dimensionless parameters  $R = r/\Omega$ and $\alpha = J/\Omega $ defined in Eq.~\eqref{rescaled_para} and the rescaled time $ \Omega \,t \to t$, the matrix elements in Eq.~\eqref{eq:rhoAt} read
\begin{equation}
\hat{\rho}_A(t) = \begin{pmatrix}
V(t)& 
W(t) \\
W^*(t) & 
1-V(t) 
\end{pmatrix}\, .
\label{eq:rhoAt_VW}
\end{equation}
where $V(t)$ and $W(t)$ are given by 
\begin{eqnarray}
V(t) &=& \frac{1}{2} \Big[1- \cos (\alpha  t)  \cos \left(\gamma t\right)  \nonumber- \frac{\alpha}{\gamma}  \sin (\alpha  t) \sin \left(\gamma t\right) \Big]  \nonumber \\
W(t) &=&  -\frac{\sin \left(\gamma t\right) \left(\alpha  \sin \left(\gamma t\right)+i\, \gamma \cos (\alpha  t)\right)}{2 \gamma^2}\,,
\label{eq:VW}
\end{eqnarray}
with $\gamma = \sqrt{\alpha^2+1}$. We substitute Eq.~\eqref{eq:rhoAt_VW} in Eq.~\eqref{eq:reset1_partial} and obtain
\begin{equation}
\hat{\rho}_{A,r}(t) = \begin{pmatrix}
V_r(t)& 
W_r(t) \\
W_r^*(t) & 
1-V_r(t) 
\end{pmatrix}\, ,
\label{eq:rhoAt_VW_r}
\end{equation}
where 
\begin{eqnarray}
V_r(t) &=&   e^{-Rt} V(t) +R \int_0^t d\tau\, e^{-R\tau} V(\tau) \nonumber \\
W_r(t) &=&    e^{-Rt} W(t) +R \int_0^t d\tau\, e^{-R\tau} W(\tau),
\label{eq:VW_r}
\end{eqnarray}
with $V(t)$ and $W(t)$ given in Eq.~\eqref{eq:VW}. Evaluating the integrals in Eq.~\eqref{eq:VW_r} one gets $V_r(t)$ and  $W_r(t)$. These expressions are a bit too long and hence they are provided in Appendix~\ref{app_ee}. We now need to compute the two eigenvalues  $\lambda_1(t)$ and  $\lambda_2(t)$ of $\hat{\rho}_{A,r}(t)$ in Eq.~\eqref{eq:rhoAt_VW_r}. Clearly $\lambda_2(t) = 1 - \lambda_1(t)$. Moreover, their product 
\begin{equation}
\lambda_1(t)\lambda_2(t) = \rm{det}[{\hat{\rho}_{A,r}(t)}] = V_r(t) (1-V_r(t))-|W_r(t)|^2 \, . 
\label{eq:lamlam}
\end{equation}
Hence, the eigenvalues are given by 
\begin{equation}
\lambda_{1,2}(t) = \frac{1\pm \sqrt{1-4  \,\rm{det}[\hat{\rho}_{A,r}(t)}}{2}\, .
\label{eq:l1t}
\end{equation} 
Consequently, the von Neumann entropy in Eq.~\eqref{eq:vnent} can be expressed as  
\begin{equation}
S_r(t) =   \ln(2)-\frac{(1+y)}{2}\ln(1+y)- \frac{(1-y)}{2}\ln(1-y)\,.
\label{eq:SY}
\end{equation}
where $y = \sqrt{1- 4\, \rm{det} [\hat{\rho}_{A,r}(t)]}$. Using the explicit expressions of $V_r(t)$ and $W_r(t)$ in Appendix~\ref{app_ee}, Eq.~\eqref{eq:SY} gives us the von Neumann entropy exactly for all $t$.

In the steady state ($t \to \infty$ limit), the matrix elements $V_r(t) \to V_r(\infty)$ and $W_r(t) \to W_r(\infty)$ whose expressions are given by
\begin{eqnarray}
\label{eq:vrinf}
V_r(\infty) &=&  \frac{1+R^2}{2+2R^2(2+R^2+4\alpha^2)}\\
W_r(\infty) &=&   -\frac{\alpha}{R^2+4+4\alpha^2} -  i \frac{R+R^3}{2+2R^2(2+R^2+4\alpha^2)}\, . \nonumber \\ \, .
\label{eq:wrinf}
\end{eqnarray}
In this case, the reduced density matrix in the NESS $\hat{\rho}_{A,r}(\infty)$ is given by the $2\times 2$ matrix
\begin{equation}
\hat{\rho}_{A,r}(\infty) = \begin{pmatrix}
V_r(\infty)& 
W_r(\infty) \\
W_r^*(\infty) & 
1-V_r(\infty) 
\end{pmatrix} \, ,
\label{eq:rhoAt_VW_r_main}
\end{equation}
where $V_r(\infty)$ and $W_r(\infty)$ are given respectively in Eqs.~\eqref{eq:vrinf} and \eqref{eq:wrinf}. 
Consequently, the von Neumann entropy in the NESS $S_r(\infty)\equiv S_{\rm st} (R,\alpha)$ can be determined explicitly 
by taking the $t\to \infty$ limit in Eq. (\ref{eq:SY}), i.e.,
\begin{eqnarray}
S_{\rm st} (R,\alpha) =   \ln(2) &-& \frac{(1+y_\infty)}{2}\ln(1+y_\infty) \nonumber \\ &-& \frac{(1-y_\infty)}{2}\ln(1-y_\infty)\,.
\label{eq:srinfty}
\end{eqnarray}
with
\begin{equation}
y_{\infty} =  \sqrt{1- 4\, \rm{det} [\hat{\rho}_{A,r}(\infty)]}\, .
\label{eq:yinfi}
\end{equation}
We next analyse this
NESS entropy $S(R, \alpha)$ in various regions of the $(R, \alpha)$ plane.

\vskip 0.5cm
\noindent (i) \textit{Noninteracting limit ($\alpha \to 0$): } For the noninteracting spins in the absence of resetting ($R=0$), the von Neumann entropy at any finite time $t$ is strictly zero, since the state of the full system remains factorized at all times $t$. However, if one switches on a finite resetting rate $R$ to the $\mid \downarrow \downarrow \rangle$ state, it has two consequences: (a) it induces strong `classical' correlations between the spins at any finite time $t$ even though there is no direct interaction between them in the Hamiltonian, similar to the results in Ref.~\cite{MCPL22} and (b) it drives the system into a NESS with strong classical
correlations.
In this NESS, the von Neumann entropy $S_{\rm st} (R,0)$ is given by a compact expression (see Appendix~\ref{app_ee})
\begin{eqnarray}
S_{\rm st}(R,0) &=& \ln(2) + \frac{1}{2}\ln(1+R^2)\nonumber \\ &+& \frac{R}{2\sqrt{R^2+1}}\ln\left(\frac{\sqrt{R^2+1}-R}{\sqrt{R^2+1}+R}\right)\, .
\label{SbarR0}
\end{eqnarray}

\begin{figure}
\includegraphics[width=0.9\linewidth]{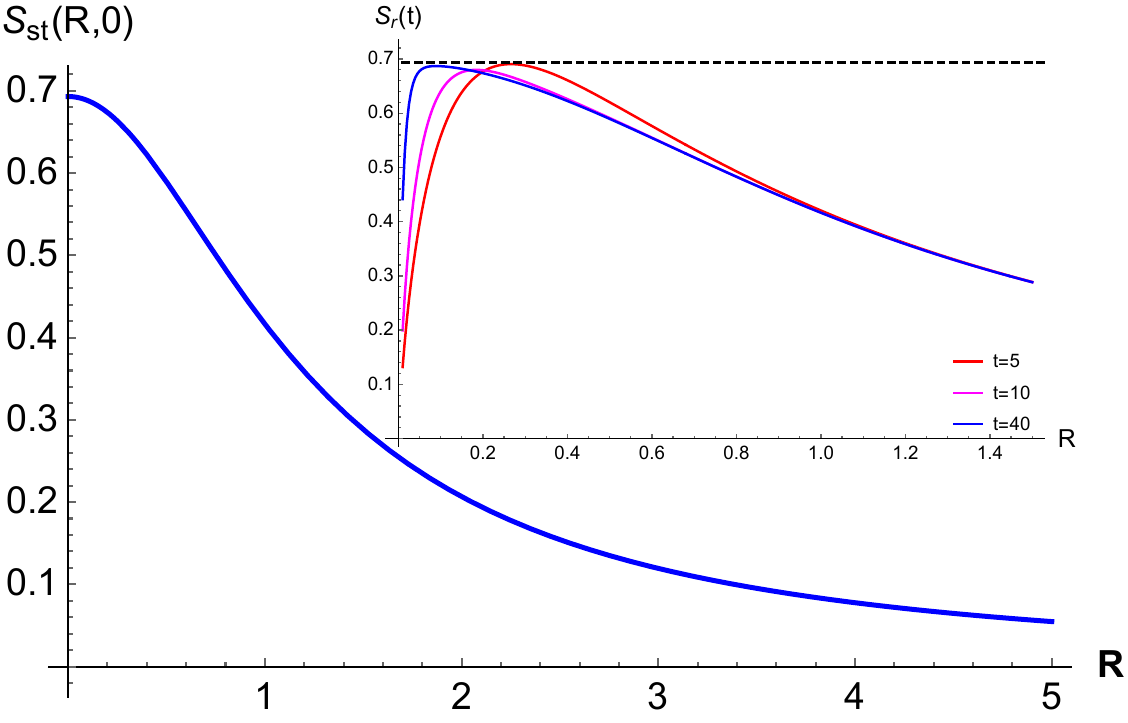}
\caption{The steady state von Neumann entropy $S_{\rm st}(R,0) $ in Eq.~\eqref{SbarR0} is plotted vs. $R$. It has the maximal value $\ln2$ at $R\to 0^+$. The inset shows the time dependent entropy $S_r(t)$ in Eq.~\eqref{eq:SY} vs. $R$ for three values of time $t$. For a finite $t$, the entropy $S_r(t)$ rises sharply from $0$ as $R\to 0^+$ to a maximum at $R=R^*(t)$, before decreasing monotonically with $R$ for $R>R^*(t)$. As $t$ increases, $R^*(t)$ decreases to zero. Eventually in the steady state ($t \to \infty$), the early time growing regime disappears leading to a monotonically decreasing entropy as a function of $R$. The horizontal line in the inset denotes the maximal value $\ln 2$ of the entropy.}
\label{fig:SRO}
\end{figure}

As discussed earlier, the von Neumann entropy in the NESS in Eq. (\ref{SbarR0}) is not a useful measure of the quantum 
entanglement since the NESS density matrix is mixed. This is evident in this noninteracting limit where the quantum entanglement measured by
concurrence is identically zero, while the von Neumann entropy in Eq.~(\ref{SbarR0}) is clearly nonzero. Thus, in the
noninteracting limit, the von Neumann entropy $S_{\rm st}(R,0)$ contains information only about classical correlations
between the spins induced by resetting.
 The entropy $S_{\rm st}(R,0)$  in Eq.~\eqref{SbarR0} decreases monotonically with increasing $R$ with a maximum $S_{\rm st}(0,0) = \ln(2)$ at $R\to0^+$, as seen in Fig.~\ref{fig:SRO}. Note that $\ln(2)$ is the maximum possible von Neumann entropy achievable in a two spin system. Thus, even in the non-interacting case, the resetting induces finite classical correlations between the spins
for any  finite $R$ and in particular even in the $R \to 0^+$ limit where the von Neumann entropy takes the maximal allowed value $\ln 2$.  This result may look a bit surprising at first sight because one expects that the entropy in a non-interacting system should vanish when $R\to 0^+$. This is of course true at any finite time $t$. However, if one takes the $t \to \infty$ limit first keeping $R$ finite, the system is driven to a NESS with
nontrivial resetting-induced correlations. Subsequently, if one takes the $R\to 0^+$ limit, the entropy remains finite in the $R\to 0^+$ limit. This is a direct consequence of the fact that the two limits $\lim_{t\to \infty} \lim_{r\to 0}$ and $\lim_{r\to 0} \lim_{t\to \infty}$ do not commute.  In fact, this is clearly seen in the time dependent behaviour of the entropy $S_r(t)$ in Eq.~\eqref{eq:SY}. At any finite time $t$, as one increases $R$, the entropy $S_r(t)$ rises sharply from its value $0$ at $R\to0^+$, achieves a maximum at $R=R^*(t)$ and then decreases with increasing $R$. As time increases, the location of the maximum $R^*(t)$ approaches zero, i.e., the maximum gets shifted towards $R=0$. Finally in the steady state, the small $R$ regime where the entropy increases sharply with $R$ shrinks to zero. This is shown in the inset of Fig.~\ref{fig:SRO} where we plot $S_r(t)$ vs. $R$
for three different times.  One may also wonder why the two noninteracting spins get maximally correlated in the NESS in the
$R\to 0^+$ limit. This is due to the fact that in the zero resetting limit (after the system reaches the NESS), the reduced
density matrix of spin 1 becomes diagonal with equal probability $1/2$
 to be in the up or in the down state, as shown in Appendix~\ref{app_ee}.  In the noninteracting case, the spin 1 has an up-down symmetry in the absence of resetting. Thus, the role of $R\to 0^+$ limit is just to ensure that
 the system reaches a NESS, but a vanishing resetting rate does not break the up down symmetry of spin 1, thus leading
 to equal probability for the up and down state for spin 1 in the NESS.

 (ii) \textit{Vanishing resetting limit in the interacting case ($\alpha > 0$):} We have seen above that a vanishing resetting rate $R\to 0^+$ drives a pair of non-interacting spins to a NESS where the von Neumann entropy saturates to its maximally allowed value $\ln 2$. A natural question is what the interaction does to this von Neumann entropy in the NESS when $R\to 0^+$. For a non-zero interaction strength $\alpha>0$, taking the  $R\to 0^+$ limit in the general expression of the reduced density matrix [see Eq.~\eqref{eq:rhoAt_VW_r_ss}], one finds 
 \begin{equation}
\hat{\rho}_{A,r}(\infty)\Big|_{R\to 0^+} = \begin{pmatrix}
1/2& 
   -\frac{\alpha}{4(\alpha^2+1)} \\
   -\frac{\alpha}{4(\alpha^2+1)} & 
1/2
\end{pmatrix}  \, .
\label{eq:rhoAt_ag0}
\end{equation}
 Thus, the presence of the interaction makes the off-diagonal elements non-zero in the NESS. Consequently, the von Neumann entropy from Eq.~\eqref{eq:srinfty} is given by 
\begin{eqnarray}
S_{\rm st}(0,\alpha) &=& \ln(2) - \frac{1}{2}\left(1+\frac{\alpha}{2(1+\alpha^2)} \right) \ln \left(1+\frac{\alpha}{2(1+\alpha^2)} \right) \nonumber \\ &-& \frac{1}{2}\left(1-\frac{\alpha}{2(1+\alpha^2)} \right) \ln \left(1-\frac{\alpha}{2(1+\alpha^2)} \right) \, .
\label{Sbarag0}
\end{eqnarray} 
 
\begin{figure}
\includegraphics[width=.95\linewidth]{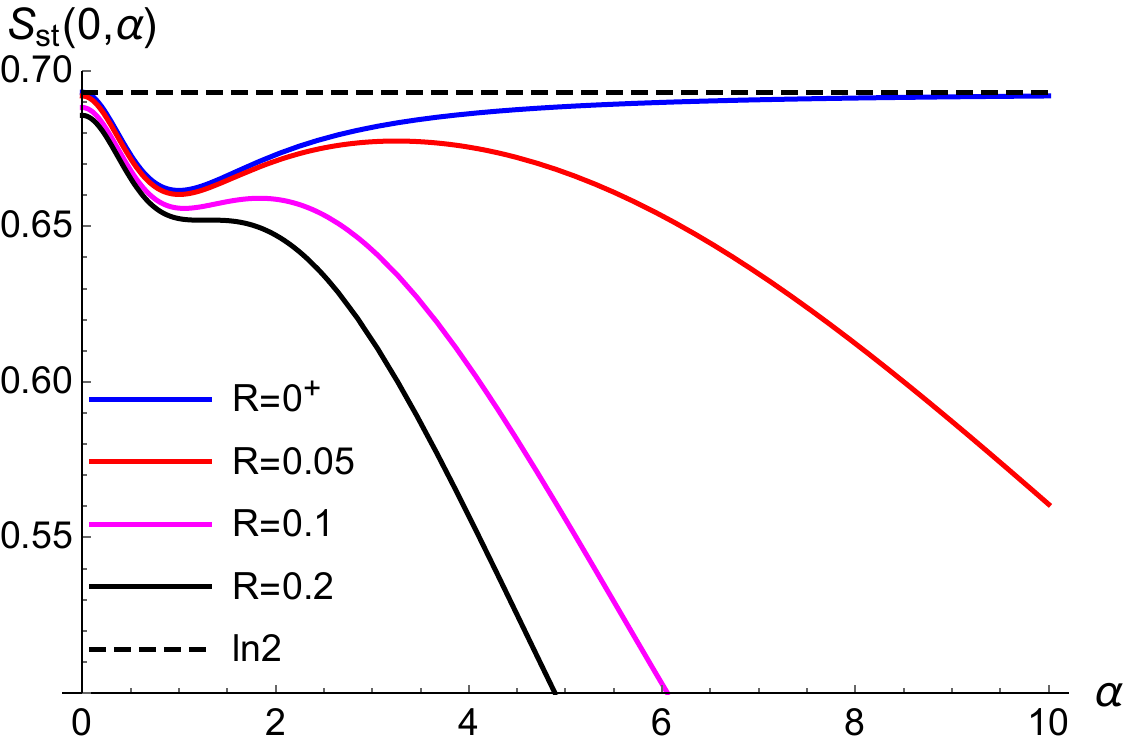}
\caption{The von Neumann entropy $S_{\rm st}(R,\alpha)$ in Eq.~\eqref{eq:srinfty} is plotted as a function of $\alpha$ for different values of $R$. In the limit $R\to 0 $, the entropy achieves its maximum value $\ln 2$ in both limits $\alpha\to0$ and $\alpha \to\infty$ with a dip in between. For small $R$, this curve has a minimum followed by a maximum beyond which the entropy decreases monotonically to zero as $\alpha \to \infty$. Finally, at a critical value $R_c\approx 0.12$, the maximum and the minimum coincide giving rise to an inflection point at $\alpha_c\approx 1.27$. For $R>R_c$,  the entropy becomes a monotonically decreasing function of $\alpha$. The transition at $R=R_c$ is reminescent of a spinodal transition. }
\label{fig:SRO1}
\end{figure}

A plot of Eq.~\eqref{Sbarag0} is shown in Fig.~\eqref{fig:SRO1} where one sees that the entropy is a non-monotonic function of $\alpha$. It achieves the maximum value $\ln 2$ in the two limits $\alpha \to 0 $ and $\alpha \to \infty$, with a dip at $\alpha = \alpha_c$. It approaches the limiting values as

\begin{equation}
S_{\rm st}(0,\alpha) \to\begin{cases} \ln 2 -\frac{\alpha^2}{8} \,\,\,\,\,\quad {\rm as} \,\, \alpha \to 0 \\ 
\ln 2  - \frac{1}{8\alpha^2} \,\,\,\quad  {\rm as} \,\,  \alpha \to \infty \, .
\end{cases}
\end{equation}
Indeed, from Eq.~\eqref{eq:rhoAt_ag0}, one sees that the off-diagonal elements vanish in both limits $\alpha \to 0$ and $\alpha \to \infty$, leading to the maximum entropy. The maximal entropy in the non-interacting limit ($\alpha \to 0$) 
has been discussed earlier in the paper. In the strongly interacting limit  ($\alpha \to \infty$) the pair of spins behave as a single `dimer' and in the $R \to 0^+$ limit, one arrives at a NESS where the spin 1 in the dimer still has equal probability to be in the up or in the down state. This also then leads to maximum von-Neumann entropy.

One can ask what happens to the von Neumann entropy as a function of $\alpha$ as one increases the resetting rate $R$. For general $R$, the exact steady state von Neumann entropy is given in Eq.~\eqref{eq:srinfty}. 
In Fig.~\ref{fig:SRO1}, we plot $S_{\rm st} (R,\alpha)$ vs. $\alpha$ for various values of $R$. As seen in the figure, for any finite $R$, the entropy vanishes as $\alpha \to \infty$. This is expected because in the strongly interacting case, finite $R$ drives the system into a NESS where the $\mid \downarrow \downarrow\rangle$ state occurs with probability $1$ since the reduced density matrix in Eq.~\eqref{eq:rhoAt_VW_r_main} approaches to $\{ \{0, 0 \}, \{ 0,1\}\} $.  
Since the $\mid \downarrow \downarrow\rangle$ state is fully factorized (i..e., completely unentangled) the von Neumann entropy vanishes. 
As $R$ increases from $0$ to a small value, the entropy $S_{\rm st} (R,\alpha)$, as a function of $\alpha$ displays a
nonmonotonic behaviour: it decreases to a minimum, then increases to a maximum and finally decreases monotonically to $0$ (algebraically)  as $\alpha\to \infty$ (see Fig.~\ref{fig:SRO1}). With increasing $R$, the height of the maximum decreases and finally at a critical value $R_c$,
the minimum and the maximum merge forming an inflection point, reminiscent of a spinodal phase transition~\cite{KB87}. 
This inflection point $(R_c,\alpha_c)$ can be obtained by setting $dS_{\rm st} (R,\alpha)/d\alpha = 0$ and $d^2 S_{\rm st} (R,\alpha)/d\alpha^2 = 0$ which gives $(R_c,\alpha_c)\approx(0.12,1.27)$ . For $R>R_c$, the entropy decreases monotonically with increasing $\alpha$.  Finally, in the strongly interacting limit $\alpha \gg 1$, as one increases $R$ infinitesimally, the entropy crosses over from the maximal value $\ln 2$ to zero. This crossover, in the limit $\alpha\to\infty$ and $R\to0^+$, is captured nicely via the scaling form $S_{\rm st} (R,\alpha) \approx F(\alpha R)$. The scaling function $F(z)$ can be computed explicitly (see Appendix~\ref{app_ee}) with the asymptotic behaviours: 
\begin{equation}
F(z)  \to  \begin{cases} \ln 2 - 8 z^4 \,\, \, \quad{\rm as}\,\, z\to0 \\
 \frac{1}{4z^2}\, \ln z \quad \,\, \quad {\rm as}\,\, z\to \infty. 
\end{cases}
\label{fzmain}
\end{equation}
Thus the entropy decreases extremely slowly as a power law (with logarithmic correction) as the interaction strength $\alpha$ increases. 
\noindent

 \begin{figure}
\includegraphics[width=0.9\linewidth]{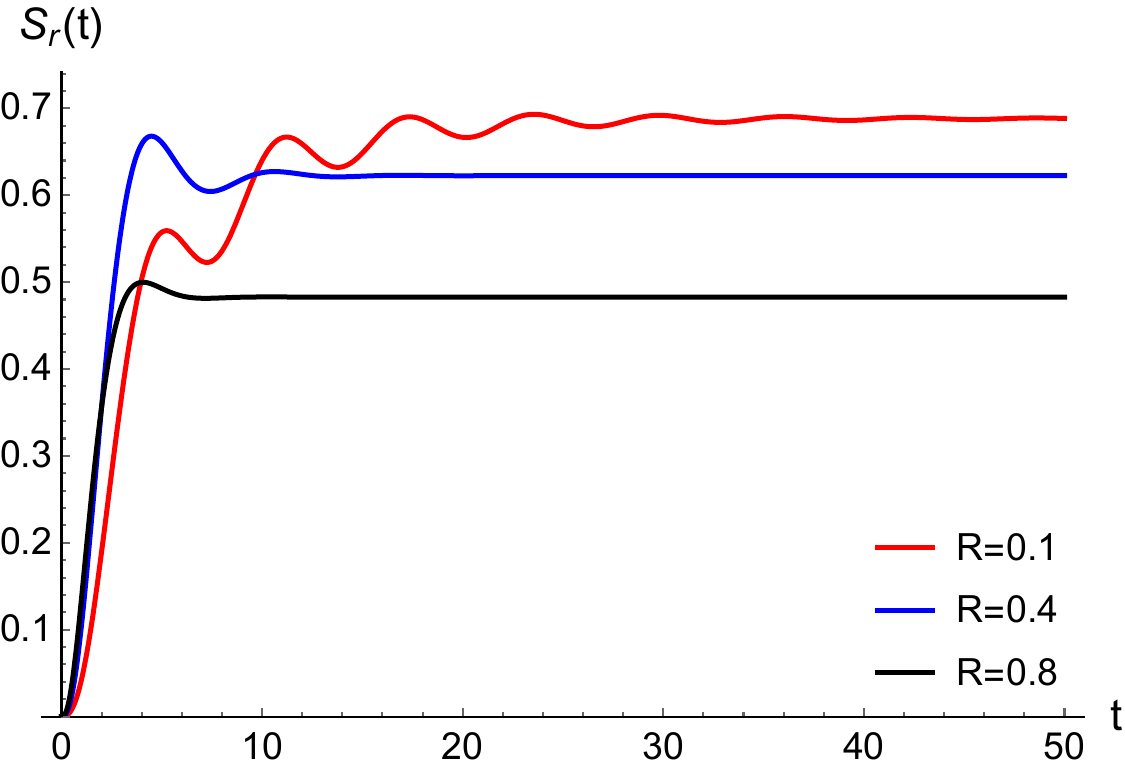}
\caption{The temporal growth of the von Neumann entropy $S_r(t)$ vs. $t$ for different values of $R$ in the noninteracting case $\alpha =0$. For finite $R$, the entropy approaches its stationary value $S_{\rm st} (R, 0)$, given in Eq.~\eqref{SbarR0}, exponentially fast with pronounced oscillations (for small $R$). Similar behaviour is also observed for the interacting case ($\alpha>0$).}
\label{fig:f3}
\end{figure}

Let us point out that unlike in the noninteracting limit where a nonzero von Neumann entropy in the NESS is solely due to the classical correlations, the situation is different in the interacting case.
Here a nonzero von Neumann entropy in the NESS has its origin in both classical and quantum correlations and it is hard to
separate their contributions. Thus to detect the purely quantum correlations, i.e., the entanglement one needs to go beyond
the von neumann entropy and study, for instance, the concurrence which will be computed in Section~\ref{conc}. 

\vskip 0.3cm
\noindent 
\textit{Approach to the steady state:} So far we have discussed the von Neumann entropy in the NESS. However, our exact result gives access to the entropy explicitly at all times and not just in the stationary limit. For simplicity, we focus on the non-interacting case ($\alpha=0$) where the time dependent entropy $S_r(t)$ takes a simpler form (see Appendix~\ref{app_ee}) and is plotted vs. $t$ for various values of $R$ in Fig.~\eqref{fig:f3}. We see that as $t\to\infty$, the entropy $S_r(t)$ approaches its steady state value $S_{\rm st}(R,0)$ given in Eq.~\eqref{SbarR0} exponentially fast (with oscillations
that are prominent for small $R$). 
Thus one  sees that a finite resetting induces a rich temporal dynamics of the von Neumann entropy.

\subsection{ Fidelity}
\label{fid}

In this two spin model, it is also natural to ask:
How far is the steady state (represented by a mixed density matrix) from the initial state represented by
a pure density matrix $\mid \downarrow\downarrow \rangle \langle  \downarrow\downarrow \mid$ ? This can be measured via the fidelity $\mathcal{F}(\hat{\rho}_r(\infty), \hat{\rho}_r(0))$ define in Eq.~\eqref{fid_def_simp1}. Using $\psi(0)=\, \mid\downarrow\downarrow\rangle$ in Eq.~\eqref{fid_def_simp1}, the fidelity is given by the diagonal matrix element of  $4 \times 4$ NESS density matrix
\begin{equation}
\mathcal{F}(\hat{\rho}_r(\infty), \hat{\rho}_r(0)) \equiv \mathcal{F}_{\rm st} (R,\alpha) =  \langle \downarrow\downarrow | \hat{\rho}_r(\infty) | \downarrow\downarrow \rangle\, ,
\end{equation}
where $\hat{\rho}_r(\infty)$ is given in Eq.~\eqref{eq:reset1_ness}. Evaluating this matrix element yields an explicit expression for the fidelity 
\begin{eqnarray}
\mathcal{F}_{\rm st} (R,\alpha) = 1 &-&\frac{1}{2}\frac{R^2+1}{\big(1+R^4+R^2\left(4 \alpha ^2+2\right)\big)} \nonumber \\ &-& \frac{1}{2}\frac{1}{(4 \alpha ^2+R^2+4)}
\label{fid:tls}
\end{eqnarray}
It is easy to check that the fidelity in Eq.~\eqref{fid:tls} lies between $0$ and $1$ everywhere in the $(R,\alpha)$ plane. When its value  is close to $1$ it indicates that the final density matrix is close to the initial one, while when the fidelity vanishes the steady state density matrix is farthest from the initial one. 

\begin{figure}
\includegraphics[width=0.9\linewidth]{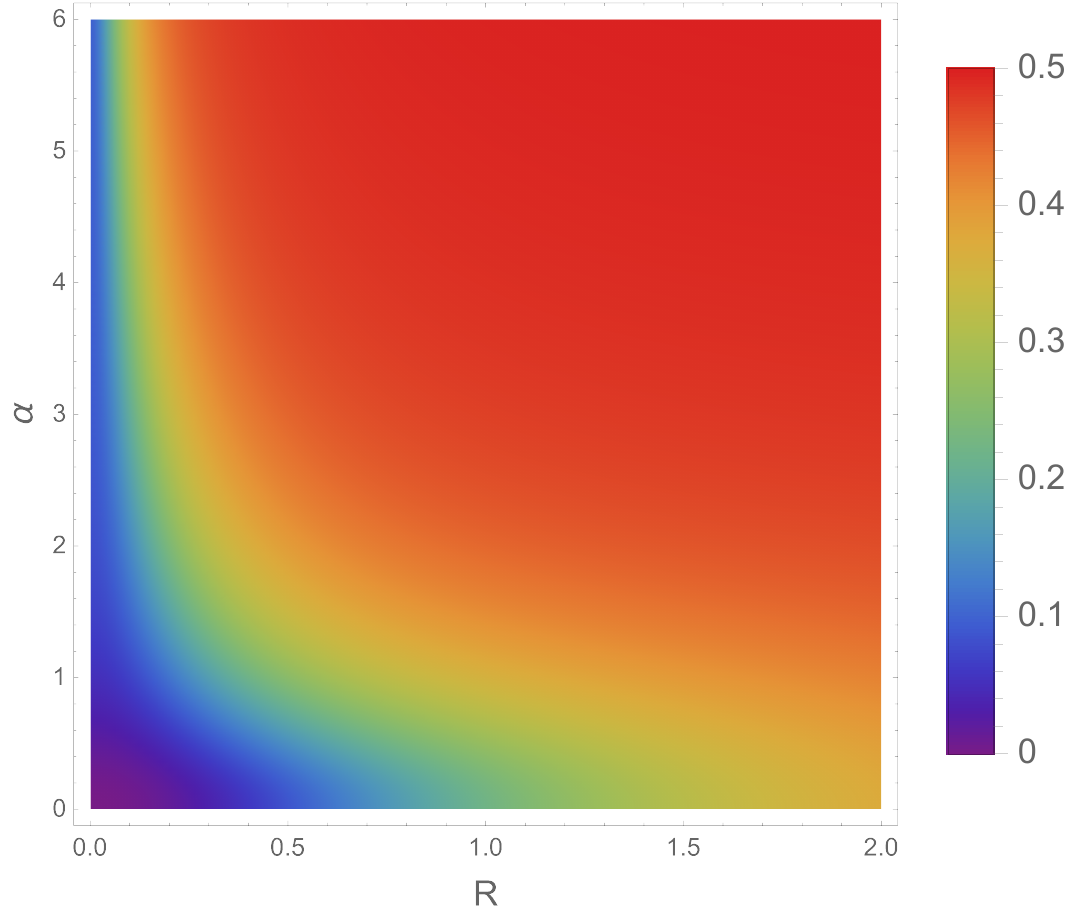}
\caption{Heatmap of the fidelity $\mathcal{F}_{\rm st} (R,\alpha)$ in Eq.~\eqref{fid:tls} shown in the $(R,\alpha)$ plane. As seen clearly in this figure, the fidelity increases monotonically with increasing $R$ or increasing $\alpha$. The colorbar on the right  indicates the magnitude of the fidelity, increasing from blue to red.}
\label{fig:fid}
\end{figure}

In Fig.~\ref{fig:fid}, we provide a heat map of the fidelity in the  $(R,\alpha)$ plane. We see that the fidelity increases monotonically when either $\alpha$ or $R$ increases and approaches to $1$ when $\alpha \to\infty$ or $R\to \infty$. This is easy to understand since in either of these limits the system is driven to the  $\mid \downarrow\downarrow\rangle$ state and consequently
the NESS density matrix is fully fidel to the initial state. Let us further discuss the two limiting cases: 
\begin{enumerate}
\item The noninteracting limit $\alpha \to 0$: In this case Eq.~\eqref{fid:tls} reduces to 
\begin{equation}
\mathcal{F}_{\rm st} (R,0) = 1 - \frac{1}{2\,(4+R^2)} - \frac{1}{2(1+R^2)}\, .
\label{fidR0}
\end{equation}
It has the following limiting behaviours for small and large $R$ 
\begin{equation}
\mathcal{F}_{\rm st} (R,0) \begin{cases} \frac{3}{8} + \frac{17}{32}R^2 + O(R^4)\quad \rm{as}\,\,\, R\to0^+\\
1 - \frac{1}{R^2} + O(R^{-4})\,\,\, \quad \rm{as}\,\,\, R\to\infty \, .
\end{cases}
\end{equation}
Interestingly, in the $R\to 0^+$ limit, the fidelity approaches the value $3/8$ which is less than unity, implying that
the steady state density matrix is far from the initial density. 

\item Vanishing resetting limit $R\to 0^+$: In this case, Eq.~\eqref{fid:tls} reduces to 
\begin{equation}
\mathcal{F}_{\rm st} (0^+,\alpha) =  \frac{3+4\alpha^2}{8(1+\alpha^2)} \, .
\label{fidvR}
\end{equation}
The asymptotic behaviours are given by 
\begin{equation}
\mathcal{F}_{\rm st} (0^+,\alpha)  = \begin{cases} \frac{3}{8} + \frac{\alpha^2}{8}+ O(\alpha^4)\,\,\quad \quad \rm{as}\,\,\, \alpha \to0^+\\
1 - \frac{1}{8\alpha^2} + O(\alpha^{-4})\,\,\, \quad \rm{as}\,\,\, \alpha\to\infty \, .
\end{cases}
\end{equation}
In the strongly interacting limit, $\alpha \to \infty$, the system has equal probability to be in the $\mid\uparrow \uparrow \rangle$ and $\mid\downarrow \downarrow \rangle$ state -- as reflected by the limiting value $1/2$ of the fidelity. 
\end{enumerate}

We also computed the purity of the NESS density matrix $\rm{tr}\big[\hat{\rho}_r(\infty)^2 \big]$ where $\hat{\rho}_r(\infty)$ is given in Eq.~\eqref{eq:reset1_ness}. Interestingly, we found that it is identical to the fidelity $\mathcal{F}_{\rm st} (R,\alpha) $ for all $R$ and $\alpha$. There is a priori no reason that the fidelity coincides with the purity in this resetting NESS, and proving this relationship
remains interesting.

\subsection{Concurrence}
\label{conc}

\begin{figure}
\includegraphics[width=0.9\linewidth]{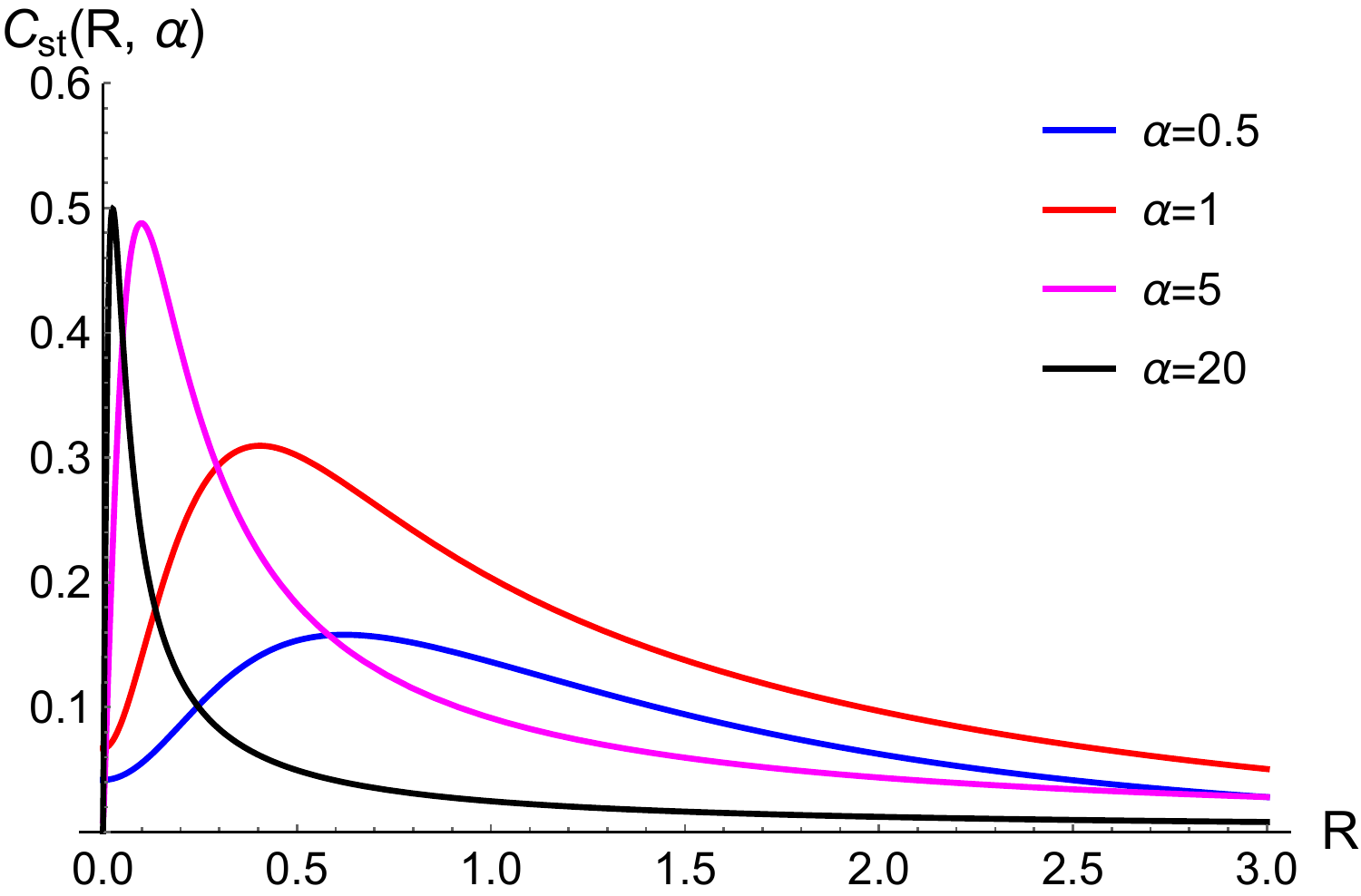}
\caption{The concurrence $\mathcal{C}_{\rm st}(R,\alpha)$ defined in Eq.~\eqref{eq:C1_tls} as a function of $R$, for different values of the interaction strength $\alpha$. For any given $\alpha$, the concurrence, as a function of $R$ increases from a nonzero value
at $R\to 0^+$, achieves a maximum, and then decreases monotonically to $0$ as $R\to \infty$.
 The peak value of the concurrence saturates to $0.5$ as $\alpha$ increases.}
\label{fig:f4}
\end{figure}

\begin{figure}
\includegraphics[width=0.9\linewidth]{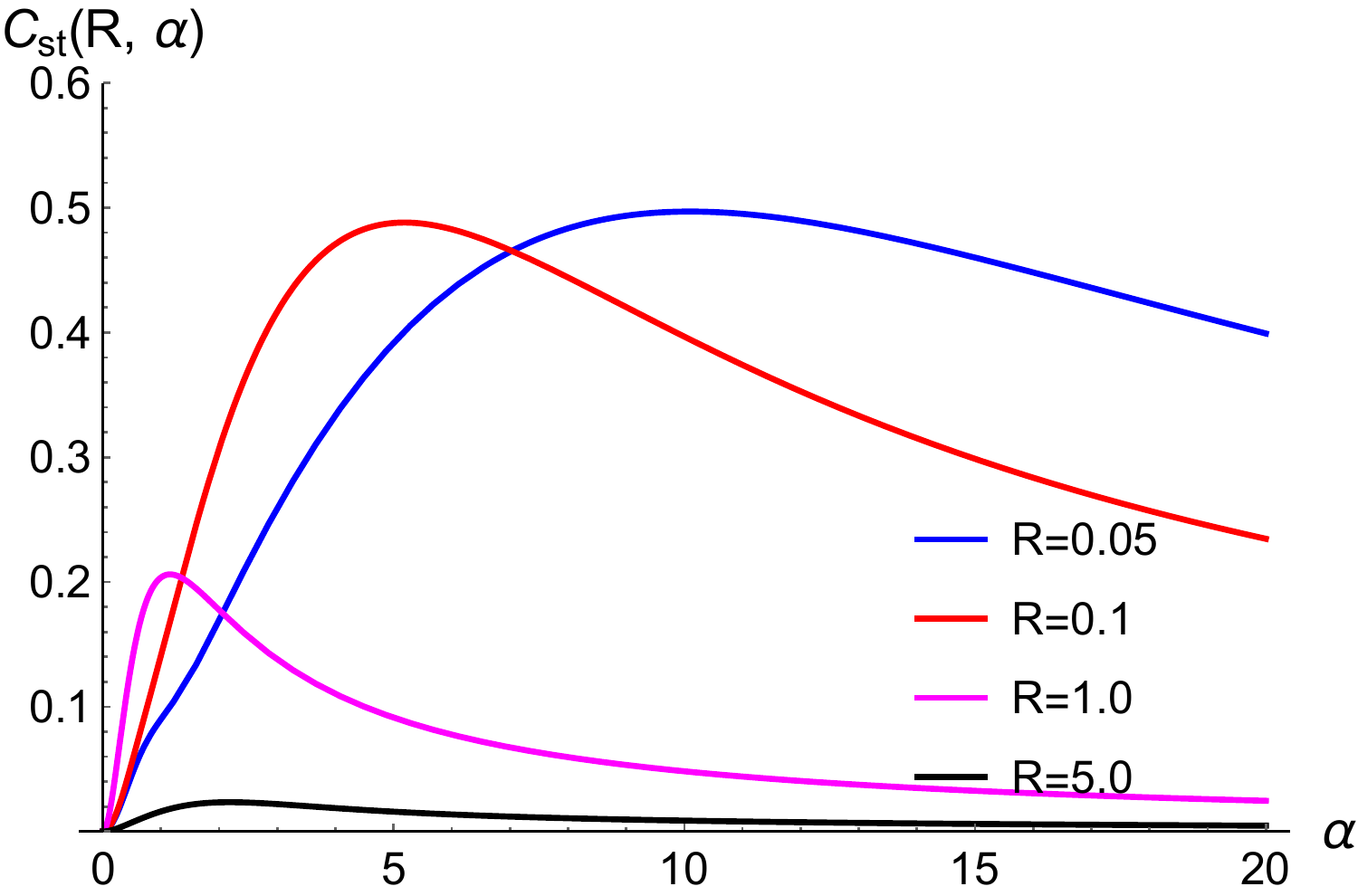}
\caption{The concurrence $\mathcal{C}_{\rm st}(R,\alpha)$ defined in Eq.~\eqref{eq:C1_tls} as a function of $\alpha$, for different values of the resetting rate $R$. For any given $R$, the concurrence, as a function of $\alpha$ first increases, achieves a
maximum, and then decays rather slowly to $0$ as $\alpha\to \infty$. The value of concurrence at the peak decreases with
increasing $R$.}
\label{fig:f5}
\end{figure}

\begin{figure}
\includegraphics[width=0.9\linewidth]{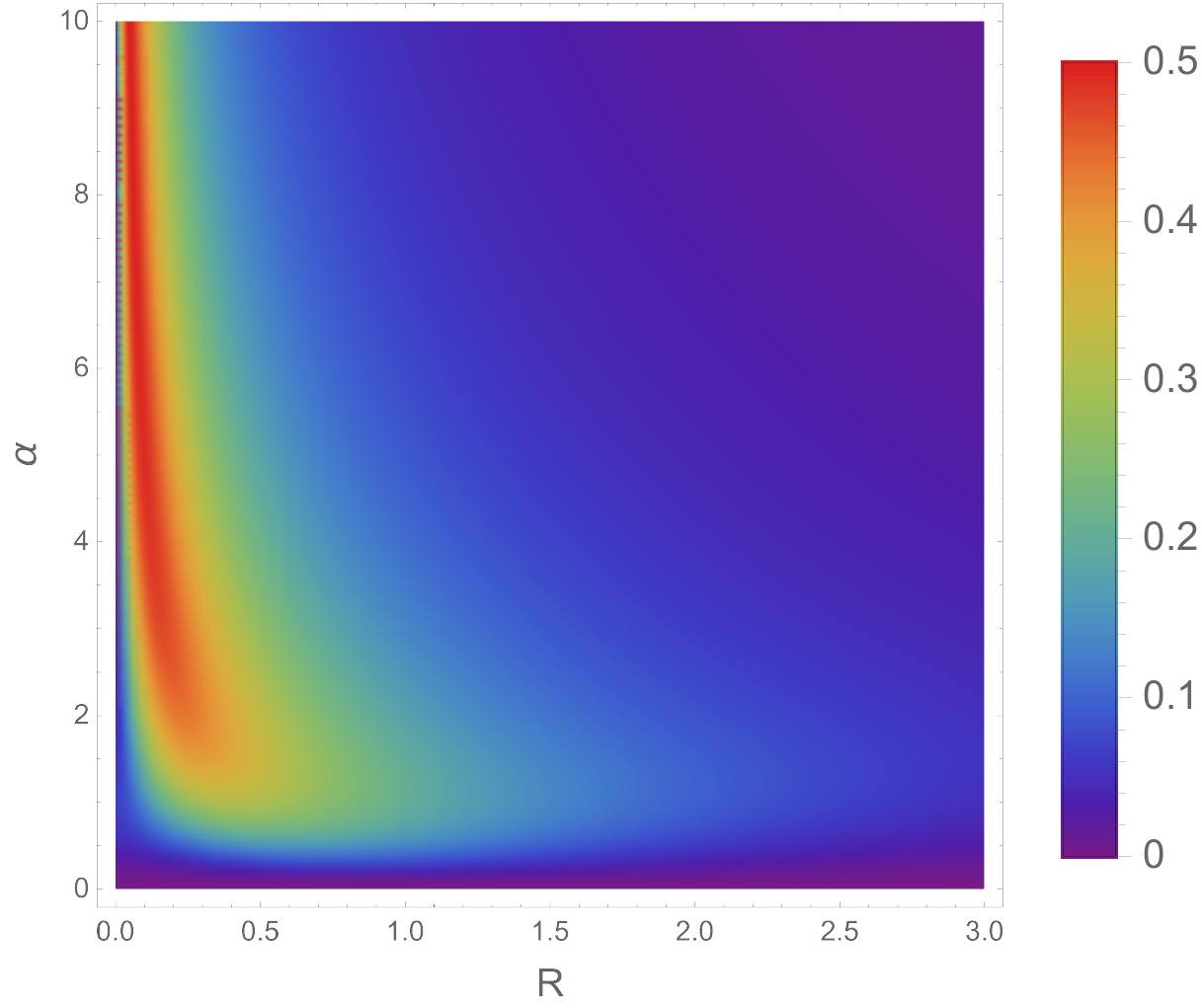}
\caption{Heatmap of the concurrence $\mathcal{C}_{\rm st}(R,\alpha)$ in Eq.~\eqref{eq:C1_tls} shown in the $(R,\alpha)$ plane. 
One sees that the concurrence vanishes for large $R$ for any interaction strength $\alpha$ and seems to have higher values
for small $R$.
The colorbar on the right  indicates the magnitude of the concurrence, increasing from blue to red.}
\label{fig:f6}
\end{figure}

We now would like to compute the quantum entanglement in the NESS density matrix given in Eq.~\eqref{eq:reset1_ness} as
a function of the two parameters $R$ and $\alpha$. 
This clearly represents a mixed state and we recall 
that the quantum entanglement in a mixed state is not captured by the von Neumann entropy.
 To isolate the entanglement from classical correlations induced by resetting one should instead investigate the concurrence as defined in Eq.~\eqref{eq:C1}. Thus, for the two spin system, we need to first compute the entries of the $4\times 4$ matrix 
\begin{equation}
\mathcal{R}_r =  \sqrt{\sqrt{\hat{\rho}_r(\infty)}\, \tilde{\rho}_r(\infty) \, \sqrt{\hat{\rho}_r(\infty)}} \, ,
\label{eq:C2t}
\end{equation}
where $\tilde{\rho}_r(\infty) $ is given by
\begin{equation}
\tilde{\rho}_r(\infty) = (\sigma_y \otimes \sigma_y) \, \hat{\rho}^*_r(\infty) \,  (\sigma_y \otimes \sigma_y) \, .
\label{eq:C3t}
\end{equation}
We recall that $\hat{\rho}^*_r(\infty)$ is the complex conjugate of $\hat{\rho}_r(\infty)$. Next, we need to 
compute the eigenvalues $\mu_1,\mu_2,\mu_3$ and $\mu_4$  of the matrix $\mathcal{R}_r$ in decreasing order and finally use Eq. (\ref{eq:C1}) to compute the concurrence, i.e., 
\begin{equation}
\mathcal{C}_{\rm st}(R,\alpha) = \rm{max} \,(0, \mu_1 - \mu_2 - \mu_3 - \mu_4)\, .
\label{eq:C1_tls}
\end{equation}
Knowing the elements of the $4\times 4$ density matrix $\hat{\rho}_r(\infty)$ explicitly, all these steps above can carried out using the Mathematica. However, the entries of the matrix $\mathcal{R}_r$ in Eq.~\eqref{eq:C2t}, though explicit, are too cumbersome to display. Consequently the eigenvalues $\mu_i$'s are rather complicated also. Hence, we evaluate them numerically and plot the concurrence for different choices of the parameters $R$ and $\alpha$. In Fig.~\ref{fig:f4}, we plot $\mathcal{C}_{\rm st}(R,\alpha) $ vs. $R$ for four different values of $\alpha$. Interestingly, the concurrence has a non-monotonic behaviour as a function of $R$ for fixed $\alpha$. As $R$ increases, it first increases, achieves a maximum and then eventually decreases to zero as $R\to\infty$.  The existence of a maximum indicates that the quantum entanglement can be maximised by choosing an optimal value of the resetting rate $R$, for every fixed $\alpha$.
The value of this maximum concurrence for a fixed $\alpha$ increases as $\alpha$ increases and saturates to $0.5$ as $\alpha \to \infty$. 
A similar nonmonotonic behaviour of $\mathcal{C}_{\rm st}(R,\alpha)$  is also seen as a function of $\alpha$ for fixed values of $R$ as shown in Fig.~\ref{fig:f5}. Thus for fixed $R$, there is also an optimal value of $\alpha$ at which the concurrence and hence the entanglement gets maximised. It is then interesting to see how the concurrence behaves in the $(R,\alpha)$ plane. We show a heat map of the concurrence in Fig.~\ref{fig:f6} which clearly shows the existence of a high concurrence region where the concurrence value is close to $0.5$. Moreover, the high concurrence region seems to be concentrated close to small resetting rate $R$. Thus, in summary,  a small nonzero resetting rate $R$ and a nonzero interaction strength $\alpha$ are both crucial to drive the pair of spins to a NESS with maximal entanglement. 

\section{Conclusions and Outlook}
\label{sec:conc}
In summary, we have provided a general framework to compute the von Neumann entropy, the fidelity and the concurrence in the non-equilibrium stationary state (NESS) induced by stochastic resetting of a closed quantum system. The density matrix in the resetting induced NESS corresponds to a mixed state with both classical and quantum correlations. The purely quantum part of the correlations, i.e., the entanglement is captured by the concurrence and not by the von Neumann entropy which is a standard measure for pure states. We then applied this general framework to compute these three observables in a simple quantum system of two ferromagnetically interacting spins, subjected to stochastic resetting with rate $r$. In this paper, we focussed on the situation where the initial state is a pure state $\mid\downarrow\downarrow \rangle$ and the system resets also to the $\mid\downarrow\downarrow \rangle$ state. However, our framework can be easily extended to other starting/resetting states. We computed exactly the density matrix of the full system in the NESS and from this, the three observables, namely (i) the von Neumann entropy of spin $1$ (ii) the fidelity
between the NESS and the initial density matrix and (iii) the concurrence in the NESS, as a function of two dimensionless
parameters $R=r/\Omega$ (the rescaled  resetting rate) and $\alpha=J/\Omega$ (the rescaled interaction strength).
One of our main conclusions is that a nonzero resetting rate $R$, together with a nonzero interaction strength $\alpha$
generates quantum entanglement in the NESS (quantified by a nonzero concurrence) and moreover this concurrence
can be optimized by appropriately choosing the two parameters $R$ and $\alpha$.

There are several future directions in which our work can be extended. Here, we focussed on small systems. It is natural to extend these studies to large systems and ask how the entanglement behaves with the system size in the presence of resetting.
It is also interesting to ask if the dependence of entanglement on the system size undergoes a phase transition as $R$ increases.
If that indeed happens, then this is somewhat similar in spirit to the measurement induced phase transition in quantum systems subjected to random projective measurements~\cite{LCF19, SRN19,GH20,ZGWGHP20,RCGG20}.
Using the general framework provided in our paper,  this interesting question can be investigated, at least numerically. 

\section{Acknowledgements} 
We warmly thank Federico Carollo and Igor Lesanovsky for useful discussions and suggestions. M.~K. would like to acknowledge support from the Project 6004-1 of the Indo-French Centre for the Promotion of  Advanced Research (IFCPAR), SERB Early Career Research Award (ECR/2018/002085), SERB Matrics Grant (MTR/2019/001101) and VAJRA faculty scheme (No. VJR/2019/000079) from the Science and Engineering Research Board (SERB), Department of Science and Technology, Government of India. M. K. acknowledges support from the Department of Atomic Energy, Government of India, under Project No. RTI4001. M.~K. thanks the hospitality of LPENS (Paris), LPTHE (Paris) and LPTMS (Paris-Saclay) during several visits. S.~N.~M. acknowledges the support from the Science and Engineering Research Board (SERB, government of India), under the VAJRA faculty scheme (No. VJR/2017/000110).
The authors would like to thank the Isaac Newton Institute for Mathematical Sciences, Cambridge, for support and hospitality during the programme New statistical physics in living matter: non equilibrium states under adaptive control where work on this paper was undertaken. This work was supported by EPSRC Grant Number EP/R014604/1.

\onecolumngrid
\medskip

\appendix 
\section{The reduced density matrix and the von Neumann entropy}
\label{app_ee}

In this Appendix, we provide some details of the computation of the reduced density matrix $\hat{\rho}_{A,r}(t)$ and the von Neumann entropy $S_r(t)$. We start by providing the explicit expressions of the matrix elements in Eq.~\eqref{eq:rhoAt_VW_r} of the main text that reads 

\begin{equation}
\hat{\rho}_{A,r}(t) = \begin{pmatrix}
V_r(t)& 
W_r(t) \\
W_r^*(t) & 
1-V_r(t) 
\end{pmatrix}\, ,
\label{eq:rhoAt_VW_r_supp}
\end{equation}
where 
\begin{eqnarray}
V_r(t) &=&   e^{-Rt} V(t) +R \int_0^t d\tau\, e^{-R\tau} V(\tau) \nonumber \\
W_r(t) &=&    e^{-Rt} W(t) +R \int_0^t d\tau\, e^{-R\tau} W(\tau),
\label{eq:VW_r_supp}
\end{eqnarray}
with $V(t)$ and $W(t)$ given by
\begin{eqnarray}
V(t) &=& \frac{1}{2} \Big[1- \cos (\alpha  t)  \cos \left(\gamma t\right)  \nonumber- \frac{\alpha}{\gamma}  \sin (\alpha  t) \sin \left(\gamma t\right) \Big]  \nonumber \\
W(t) &=&  -\frac{\sin \left(\gamma t\right) \left(\alpha  \sin \left(\gamma t\right)+i\, \gamma \cos (\alpha  t)\right)}{2 \gamma^2}\,.
\label{eq:VW_supp}
\end{eqnarray}

We recall that 
\begin{equation}
\gamma = \sqrt{\alpha^2+1}\, ,
\label{eq:gamma}
\end{equation} 
and we are working in the rescaled time $\Omega \, t \to t$. Furthermore $R$ and $\alpha$ are the dimensionless resetting rate and the interaction strength defined in Eq.~\eqref{rescaled_para} of the main text. 
Substituting Eq.~\eqref{eq:VW_supp} in Eq.~\eqref{eq:VW_r_supp} and performing the integrals we obtain
\begin{equation}
V_r(t) =   V_r(\infty) +  e^{-R t}v_r(t)
\label{eq:V_r_app}
\end{equation}
where 
\begin{eqnarray}
\label{eq:Vfull_inf}
V_r(\infty) &=&  \frac{1+R^2}{2+2R^2(2+R^2+4\alpha^2)}  \\ 
\nonumber \\
v_r(t) &=& \frac{-\cos (\alpha\,  t) \Big(\gamma  \left(R^2+1\right) \cos (\gamma \, t)+R \left(2 \gamma ^2+R^2-1\right) \sin (\gamma \, t)\Big)+\alpha  \sin (\alpha\,  t) \Big(\left(R^2-1\right) \sin (\gamma \,  t)+2 \gamma  R \cos (\gamma \, t)\Big)}{8 \gamma ^3 R^2+2 \gamma  \left(R^2-1\right)^2} \, . \nonumber \\
\label{eq:Vfull}
\end{eqnarray}
Similarly, 
\begin{equation}
W_r(t) =   W_r(\infty) +  e^{-R t}w_r(t)
\label{eq:W_r_app}
\end{equation}
where 
\begin{eqnarray}
\label{eq:Wfull_inf}
W_r(\infty) &=&   -\frac{\alpha}{R^2+4+4\alpha^2} -  i \frac{R+R^3}{2+2R^2(2+R^2+4\alpha^2)}  \\
\nonumber \\
w_r(t) &=& \frac{1}{4\gamma}\bigg[  \frac{2 \, \alpha \, R \sin (2 \gamma \, t)}{4 \gamma ^2+R^2}+\frac{4\, \alpha \, \gamma  \cos (2 \gamma \, t)}{4 \gamma ^2+R^2}+\frac{i (\alpha -\gamma )^2 \sin (t (\alpha -\gamma ))}{(\alpha -\gamma )^2+R^2}-\frac{i (\alpha +\gamma )^2 \sin (t (\alpha +\gamma ))}{(\alpha +\gamma )^2+R^2}  \nonumber \\ &-& \frac{i R (\alpha -\gamma ) \cos (t (\alpha -\gamma ))}{(\alpha -\gamma )^2+R^2} +\frac{i R (\alpha +\gamma ) \cos (t (\alpha +\gamma ))}{(\alpha +\gamma )^2+R^2} \bigg]\, . \nonumber \\
\label{eq:Wfull}
\end{eqnarray}

Next, we compute the two eigenvalues  $\lambda_1(t)$ and  $\lambda_2(t)$ of $\hat{\rho}_{A,r}(t)$ in Eq.~\eqref{eq:rhoAt_VW_r_supp}. Clearly $\lambda_2(t) = 1 - \lambda_1(t)$. Moreover, their product 
\begin{equation}
\lambda_1(t)\lambda_2(t) = \rm{det}[{\hat{\rho}_{A,r}(t)}] = V_r(t) \big(1-V_r(t)\big)-|W_r(t)|^2\, . 
\label{eq:l1l2}
\end{equation}

Hence, the eigenvalues $\lambda_{1,2}(t)$ are given by 
\begin{equation}
\lambda_{1,2}(t) = \frac{1\pm \sqrt{1-4  \rm{det}[\hat{\rho}_{A,r}(t)}}{2} \, .
\end{equation}

Consequently, the von Neumann entropy can be expressed as  

\begin{equation}
S_r(t) =    - \rm{tr} \big[ \hat{\rho}_{A,r} (t) \ln(\hat{\rho}_{A,r} (t)) \big] = -\sum_{i=1}^{N_A} \lambda_i (t) \ln(\lambda_i (t)) = \ln(2)-\frac{1}{2}(1+y)\ln(1+y)- \frac{1}{2}(1-y)\ln(1-y)\, , 
\label{eq:SY_app}
\end{equation}
where 
\begin{equation}
y = \sqrt{1- 4\, \rm{det} [\hat{\rho}_{A,r}(t)]} = \sqrt{1+4\, |W_r(t)|^2 - 4\, V_r(t)(1-V_r(t))}\, ,
\label{eq:yt_app}
\end{equation}
where $V_r(t)$ and $W_r(t)$ are given respectively in Eqs.~\eqref{eq:V_r_app} and  Eqs.~\eqref{eq:W_r_app}. 
This, then gives us the exact von Neumann entropy $S_r(t)$ at all times $t$ and for arbitrary $R$ and $\alpha$. Below, we consider different limiting cases of this entropy. 

\vskip 0.2cm 
\noindent \textit{Steady State: } In the long time limit, $t\to\infty$ the system approaches a non-equilibrium stationary state (NESS) and the entropy $S_r(t)$ approaches its stationary value $S_{\rm st}(R,\alpha)$. To compute this stationary value, we take the limit $t\to\infty$ in Eq.~\eqref{eq:rhoAt_VW_r_supp}. This gives
\begin{equation}
\hat{\rho}_{A,r}(\infty) = \begin{pmatrix}
V_r(\infty)& 
W_r(\infty) \\
W_r^*(\infty) & 
1-V_r(\infty) 
\end{pmatrix} =  \begin{pmatrix}
\frac{1+R^2}{2+2R^2(2+R^2+4\alpha^2)}& 
  -\frac{\alpha}{R^2+4+4\alpha^2} -  i \frac{R+R^3}{2+2R^2(2+R^2+4\alpha^2)} \\
  -\frac{\alpha}{R^2+4+4\alpha^2} + i \frac{R+R^3}{2+2R^2(2+R^2+4\alpha^2)} & 
1-\frac{1+R^2}{2+2R^2(2+R^2+4\alpha^2)} 
\end{pmatrix}  \, .
\label{eq:rhoAt_VW_r_ss}
\end{equation}
Consequently, the von Neumann entropy in the steady state is given by 
 \begin{equation}
S_{\rm st} (R,\alpha) =  \ln(2)-\frac{1}{2}(1+y_\infty)\ln(1+y_\infty)- \frac{1}{2}(1-y_\infty)\ln(1-y_\infty)\, ,
\label{eq:Sst_supp}
\end{equation}
where 
\begin{equation}
y_\infty  =\sqrt{1-4 \, \text{det}[\hat{\rho}_{A,r}(\infty)]}  = \sqrt{1+4\, |W_r(\infty)|^2 - 4\, V_r(\infty)(1-V_r(\infty))} \, .
\label{eq:yst_supp}
\end{equation}
where $V_r(\infty)$ and $W_r(\infty)$ are given respectively in Eqs.~\eqref{eq:Vfull_inf} and \eqref{eq:Wfull_inf}.
This gives
 \begin{eqnarray}
y_\infty & =&\sqrt{1+\frac{4 \alpha ^2}{\left(4 \alpha ^2+R^2+4\right)^2}-\frac{\left(R^2+1\right) \left(\left(8 \alpha ^2+2\right) R^2+R^4+1\right)}{\left(\left(4 \alpha ^2+2\right) R^2+R^4+1\right)^2}}\, .
\label{eq:yst_supp_full}
\end{eqnarray}
Substituting this expression of $y_\infty$ in Eq.~\eqref{eq:Sst_supp} gives the the NESS entropy $S_{\rm st} (R,\alpha)$ for arbitrary $\alpha$ and $R$. 

\vskip 0.2cm
\noindent \textit{Non-interacting limit in the steady state:}  In the non-interacting case, setting $\alpha = 0$ in Eq.~\eqref{eq:yst_supp_full} one gets
$y_\infty = R/\sqrt{1+R^2}$. Substituting this in Eq.~\eqref{eq:Sst_supp} one gets Eq.~(\ref{SbarR0}) of the main text, i.e., 
\begin{eqnarray}
S_{\rm st}(R,0) &=& \ln(2) + \frac{1}{2}\ln(1+R^2)+ \frac{R}{2\sqrt{R^2+1}}\ln\left(\frac{\sqrt{R^2+1}-R}{\sqrt{R^2+1}+R}\right)\, .
\label{SbarR0_supp}
\end{eqnarray}
By further taking the $R\to 0^+$ limit one gets $S_{\rm st}(0,0)  = \ln 2$ which is the maximal allowed entropy. To understand why one obtains the maximal entropy in this non-interacting limit, it is useful to investigate the reduced density matrix in Eq.~\eqref{eq:rhoAt_VW_r_ss} which, for $\alpha = 0$, reads 
\begin{equation}
\hat{\rho}_{A,r}(\infty)\Big|_{\alpha=0} = \begin{pmatrix}
\frac{1+R^2}{2+2R^2(2+R^2)}& 
   -  i \frac{R+R^3}{2+2R^2(2+R^2)} \\
  + i \frac{R+R^3}{2+2R^2(2+R^2)} & 
1-\frac{1+R^2}{2+2R^2(2+R^2)} 
\end{pmatrix}  \, .
\label{eq:rhoAt_VW_r_ss_a0}
\end{equation}
As discussed in the main text, in the non-interacting case ($\alpha=0$), the pair of spins remains factorized at all time $t$ in the absence of resetting. If one switches on a finite resetting rate $R$ (to the $\mid \downarrow \downarrow \rangle$ state), it has
two effects: (A) a finite rate of resetting induces strong  classical correlations between the  pair of spins even though they are noninteracting
at the level of the Hamiltonian and (B) it drives the pair of spins to a NESS where the reduced density matrix of spin 1 becomes
time-independent and is given by Eq.~(\ref{eq:rhoAt_VW_r_ss_a0}). If one now takes the $R\to 0^{+}$ limit, the reduced density matrix of spin 1 becomes  
\begin{equation}
\hat{\rho}_{A,r}(\infty)\Big|_{\alpha=0, R=0} = \begin{pmatrix}
1/2& 
  0 \\
  0 & 
1/2
\end{pmatrix}  \, ,
\label{eq:rhoAt_VW_r_ss_a0r0}
\end{equation}
indicating that in the NESS the up and down states for spin 1 are equally likely.  Consequently, the von Neumann entropy takes the maximum value $ \ln 2$ in the $R\to 0^+$ limit after the spins have reached the NESS. 

\vskip 0.2cm

\noindent \textit{Interacting case ($\alpha>0$) in the steady state:}
In this case, the reduced density matrix is given in Eq.~\eqref{eq:rhoAt_VW_r_ss} for arbitrary $R$ and the von Neumann entropy in the steady state is given by Eqs.~\eqref{eq:Sst_supp}-\eqref{eq:yst_supp_full}. In the limit $R\to 0^+$ the von Neumann entropy $S_{\rm st}(0,\alpha)$ approaches the maximum value $\log 2$ as $\alpha \to\infty$ and $\alpha \to 0$ with a dip in between (see Fig~\ref{fig:SRO1} in the main text). When $R$ is small but finite, the entropy vanishes as $\alpha \to \infty$ because a finite resetting rate and strong interaction drives the system into the dimer state $\mid \downarrow \downarrow \rangle$ which is factorizable and hence is unentangled. It is then natural to ask how the entropy crosses over from its maximum value $\ln2$ to $0$ as $R$ increases slightly from $0$ in the $\alpha \to \infty$ limit. To investigate this crossover, we consider the entropy $S_{\rm st}(R,\alpha)$ given in Eqs.~\eqref{eq:Sst_supp}-\eqref{eq:yst_supp_full} in the limit $R\to0$ and $\alpha\to\infty$ limit. It turns out that if one takes these two limits simultaneously keeping the scaling combination $z=\alpha\, R$ fixed, the entropy admits a scaling form
\begin{equation}
S_{\rm st}(R,\alpha) \to F(\alpha R)\, ,
\label{eq:FaR}
\end{equation} 
where the scaling function $F(z)$ is given explicitly by 
\begin{equation}
F(z) = \ln 2-\frac{1}{2}\frac{1+8z^2}{1+4z^2}\ln \left(\frac{1+8z^2}{1+4z^2}\right)+\frac{1}{2}\frac{1}{1+4z^2}\ln \left(1+4z^2\right)\, .
\label{eq:FaR_z}
\end{equation} 
This scaling function has the asymptotic behaviour [as mentioned in Eq.~\eqref{fzmain} of the main text]
\begin{equation}
F(z) = \begin{cases} \ln 2 -8 z^4 \quad \text{as} \quad z \to 0 \\ 
\\
\frac{1}{4z^2}\,\ln z \quad \quad \quad \,\,  \text{as}\quad  z \to \infty
\end{cases}
\label{eq:FaR_z_lim}
\end{equation} 
Thus, for a fixed small $R$, as $\alpha \gg 1/R$, i.e., $z\gg 1$, the entropy decreases algebraically (with a logarithmic correction) as the interaction $\alpha$ increases. 

\vskip 0.2cm
\noindent \textit{Approach to the steady state:}
As mentioned before, Eq.~\eqref{eq:SY} provides the entropy at all times $t$ and one can easily work out how it approaches its stationary value as $t \to \infty$. Here, for simplicity, we focus on the non-interacting limit ($\alpha = 0$) where the time dependent entropy takes a simpler form. Putting $\alpha = 0$, in Eqs.~\eqref{eq:V_r_app}-\eqref{eq:Vfull}, we get 
\begin{equation}
V_r(t) \big|_{\alpha=0}= \frac{1-e^{-Rt}(\cos(t)+R\sin(t))}{2(1+R^2)} \,. 
\end{equation}
Similarly, 
\begin{equation}
W_r(t) \big|_{\alpha=0}= -\frac{i \left(R-e^{-R t} (R \cos (t)-\sin (t))\right)}{2 \left(1+R^2\right)} \,. 
\end{equation}
Therefore, from Eq.~\eqref{eq:yt_app}, we get 
\begin{equation}
y = \sqrt{\frac{R^2+e^{-2 R t}+2\, R e^{-R t} \sin (t)}{R^2+1}}\, . 
\label{eq:ya0_app}
\end{equation}
Consequently, the time dependent entropy $S_r(t)$ at $\alpha=0$ is given by Eq.~\eqref{eq:SY} with $y$ in Eq.~\eqref{eq:ya0_app}. In Fig.~(3) of the main text, we plot $S_r(t)$ at $\alpha=0$ vs. $t$ for various values of $R$. We see that as $t\to\infty$, the entropy $S_r(t)$ approaches its steady state value $S_{\rm st}(R,0)$ given in Eq.~\eqref{SbarR0_supp} exponentially fast (with oscillations
that are prominent for small $R$). 
Thus one  sees that a finite resetting induces a nontrivial temporal growth of the von Neumann entropy.


\twocolumngrid


\medskip

\begin{thebibliography}{100}

\bibitem{NC_Book_2010}
M. A. Nielsen, I. Chuang, {\textit Quantum computation and quantum information}, (Cambridge University Press,
Cambridge, 2010).

\bibitem{BZ_Book_17}
I. Bengtsson and Karol Zyczkowski,
\textit{Geometry of quantum states: an introduction to quantum entanglement},  
(Cambridge University Press, Cambridge, 2017).

\bibitem{AFOV08}
L. Amico, R. Fazio, A. Osterloh, V. Vedral, 
\textit{Entanglement in many-body systems},
Rev. Mod. Phys. \textbf{80}, 517 (2008).

\bibitem{P93}
D. N. Page,
\textit{Average entropy of a subsystem},
Phys. Rev. Lett. \textbf{71}, 1291 (1993).


\bibitem{CC04}
P. Calabrese, J. Cardy,
\textit{Entanglement entropy and quantum field theory},
J. Stat. Mech. 06002 (2004).

\bibitem{CP20}
P. Calabrese, Pasquale,
\textit{Entanglement spreading in non-equilibrium integrable systems},
SciPost Phys. Lect. Notes 20 (2020).

\bibitem{CC05}
P. Calabrese, J.  Cardy,
\textit{Evolution of entanglement entropy in one-dimensional systems},
J. Stat. Mech. 04010 (2005).

\bibitem{FMPPS08}
P. Facchi, U. Marzolino, G. Parisi, S. Pascazio, and A. Scardicchio,
\textit{Phase Transitions of Bipartite Entanglement},
Phys. Rev. Lett. \textbf{101}, 050502 (2008).

\bibitem{NMV10}
C. Nadal, S. N. Majumdar, M. Vergassola,
\textit{Phase Transitions in the Distribution of Bipartite Entanglement of a Random Pure State},
Phys. Rev. Lett. \textbf{104}, 110501 (2010).

\bibitem{SNM10}
S.N. Majumdar, 
\textit{Extreme Eigenvalues of Wishart Matrices: Application to Entangled Bipartite System}, a chapter in the book \textit{Handbook of Random Matrix Theory} (ed. by G. Akemann, J. Baik and P. Di Francesco), 
(Oxford University Press, Oxford, 2011), also available at arXiv:1005.4515. 

\bibitem{NMV11}
C. Nadal, S. N. Majumdar, M. Vergassola,
\textit{Statistical Distribution of Quantum Entanglement for a Random Bipartite State},
J. Stat. Phys. \textbf{142}, 403 (2011).

\bibitem{SLRD13}
J. Schachenmayer, B. P. Lanyon, C. F. Roos, A. J. Daley,
\textit{Entanglement Growth in Quench Dynamics with Variable Range Interactions},
Phys. Rev. X \textbf{3}, 031015 (2013).

\bibitem{CDM15}
P. Calabrese, P. Le Doussal, S.N. Majumdar, 
\textit{Random matrices and entanglement entropy of trapped Fermi gases}, 
Phys. Rev. A. \textbf{91}, 012303 (2015).

\bibitem{KTLRSPG16}
A. M. Kaufman, M. E. Tai, A. Lukin, M. Rispoli, R. Schittko, P. M. Preiss, M. Greiner,
\textit{Quantum thermalization through entanglement in an isolated many-body system},
Science \textbf{353}, 794 (2016). 

\bibitem{BEJVMLZBR19}
T. Brydges, A. Elben, P. Jurcevic, B. Vermersch, C. Maier, B. P. Lanyon, P. Zoller, R. Blatt, C. F. Roos,
\textit{Probing Rényi entanglement entropy via randomized measurements},
Science \textbf{364}, 260 (2019).


\bibitem{LMS19}
B. Lacroix-A-Chez-Toine, S.N. Majumdar, and G. Schehr, 
\textit{Entanglement Entropy and Full Counting Statistics for 2d-Rotating Trapped Fermions}, 
Phys. Rev. A. \textbf{99}, 021602 (R) (2019).

\bibitem{FG21}
S. Fraenkel, M. Goldstein,
\textit{Entanglement measures in a nonequilibrium steady state: Exact results in one dimension},
SciPost Phys. \textbf{11}, 085 (2021).

\bibitem{SKCD21}
S. Scopa, Alexandre Krajenbrink, P. Calabrese, J. Dubail,
\textit{Exact entanglement growth of a one-dimensional hard-core quantum gas during a free expansion},
J. Phys. A: Math. Theor. \textbf{54} 404002 (2021).


\bibitem{CTKC22}
M. Coppola, E. Tirrito, D. Karevski, M. Collura,
\textit{Growth of entanglement entropy under local projective measurements},
Phys. Rev. B. \textbf{105}, 094303 (2022).

\bibitem{W98}
W. K. Wootters,
\textit{Entanglement of Formation of an Arbitrary State of Two Qubits},
Phys. Rev. Lett. \textbf{80}, 2245 (1998). 

\bibitem{EP07}
V. Eisler, I. Peschel, 
\textit{Evolution of entanglement after a local quench},
J. Stat. Mech. 06005 (2007).

\bibitem{C11}
J. Cardy,
\textit{Measuring Entanglement Using Quantum Quenches},
Phys. Rev. Lett. \textbf{106}, 150404  (2011).

\bibitem{AC18}
V. Alba, P. Calabrese,
\textit{Entanglement dynamics after quantum quenches in generic integrable systems},
SciPost Phys. \textbf{4}, 017 (2018).

\bibitem{A18}
A. Mitra,
\textit{Quantum Quench Dynamics},
Annu. Rev. Condens. Matter Phys. \textbf{9}, 245 (2018).

\bibitem{LCF19}
Y. Li, X. Chen, M. P. A. Fisher,
\textit{Measurement-driven entanglement transition in hybrid quantum circuits},
Phys. Rev. B \textbf{100}, 134306 (2019).

\bibitem{SRN19}
B. Skinner, J. Ruhman, A. Nahum,
\textit{Measurement-Induced Phase Transitions in the Dynamics of Entanglement},
Phys. Rev. X \textbf{9}, 031009 (2019).

\bibitem{GH20}
M. J. Gullans, D. A. Huse,
\textit{Dynamical Purification Phase Transition Induced by Quantum Measurements},
Phys. Rev. X \textbf{10}, 041020  (2020).

\bibitem{ZGWGHP20}
A. Zabalo, M. J. Gullans, J. H. Wilson, S. Gopalakrishnan, D. A. Huse, J. H. Pixley, {\it Critical properties of the measurement- induced transition in random quantum circuits}, Phys. Rev. B \textbf{101}, 060301 (2020).

\bibitem{RCGG20}
S. Roy, J. T. Chalker, I. V. Gornyi, and Y. Gefen, {\it Measurement- induced steering of quantum systems},
Phys. Rev. Research \textbf{2}, 033347 (2020).


\bibitem{NPSC12}
C. Navarrete-Benlloch, R. García-Patrón, J. H. Shapiro, N. J. Cerf,
\textit{Enhancing quantum entanglement by photon addition and subtraction},
Phys. Rev. A \textbf{86}, 012328 (2012).

\bibitem{DBVDCK13}
M.R. Delbecq, L.E. Bruhat, J.J. Viennot, S. Datta, A. Cottet, T. Kontos,
\textit{Photon mediated interaction between distant quantum dot circuits},
Nature Communications \textbf{4}, 1400 (2013).

\bibitem{AKT14}
C. Aron, M.Kulkarni and H. E. Tureci, \textit{Steady-state entanglement of spatially separated qubits via quantum bath engineering},
Phys. Rev. A \textbf{90}, 062305 (2014).

\bibitem{RSMMVEKWSS14}
N. Roch, M. E. Schwartz, F. Motzoi, C. Macklin, R. Vijay, A. W. Eddins, A. N. Korotkov, K. B. Whaley, M. Sarovar, I. Siddiqi,
\textit{Observation of Measurement-Induced Entanglement and Quantum Trajectories of Remote Superconducting Qubits},
Phys. Rev. Lett. \textbf{112}, 170501 (2014).

\bibitem{AKH16}
C. Aron, M.Kulkarni, H. E. Tureci, \textit{Photon-mediated interactions: a scalable tool to create and sustain entangled many-body states}, Phys. Rev. X \textbf{6}, 011032 (2016).

\bibitem{SMFAKTS16}
M. E. Schwartz, L. Martin, E. Flurin, C. Aron, M. Kulkarni, H. E. Tureci, I. Siddiqi, \textit{Stabilizing entanglement via symmetry-selective bath engineering in superconducting qubits}, Phys. Rev. Lett. \textbf{116}, 240503 (2016).

\bibitem{STT16}
V. Srinivasa, J. M. Taylor, C. Tahan,
\textit{Entangling distant resonant exchange qubits via circuit quantum electrodynamics},
Phys. Rev. B \textbf{94}, 205421 (2016).

\bibitem{SSHCHMMGA17}
R. Stockill, M. J. Stanley, L. Huthmacher, E. Clarke, M. Hugues, A. J. Miller, C. Matthiesen, C. Le Gall, M. Atatüre,
\textit{Phase-Tuned Entangled State Generation between Distant Spin Qubits},
Phys. Rev. Lett. \textbf{119}, 010503 (2017).

\bibitem{WWW19}
Z. Wang, W. Wu, J.Wang, 
\textit{Steady-state entanglement and coherence of two coupled qubits in equilibrium and nonequilibrium environments},
Phys. Rev. A \textbf{99}, 042320 (2019).


\bibitem{HW97}
S. A. Hill, W. K. Wootters,
\textit{Entanglement of a Pair of Quantum Bits},
Phys. Rev. Lett. \textbf{78}, 5022 (1997).

\bibitem{HHHH09}
R. Horodecki, P. Horodecki, M. Horodecki, K. Horodecki,
\textit{Quantum entanglement},
Rev. Mod. Phys. \textbf{81}, 865 (2009).

\bibitem{OZ01}
H. Ollivier, W. H. Zurek,
\textit{Quantum Discord: A Measure of the Quantumness of Correlations},
Phys. Rev. Lett. \textbf{88}, 017901 (2001).

\bibitem{HV21}
L. Henderson and V. Vedral,
\textit{Classical, quantum and total correlations},
J. Phys. A: Math. Gen. \textbf{34} 6899 (2001).

\bibitem{GG10}
P. Giorda, M. G. A. Paris,
\textit{Gaussian Quantum Discord},
Phys. Rev. Lett. \textbf{105}, 020503 (2010).

\bibitem{GTA13}
D. Girolami, T. Tufarelli, G. Adesso,
\textit{Characterizing Nonclassical Correlations via Local Quantum Uncertainty},
Phys. Rev. Lett. \textbf{110}, 240402 (2013).

\bibitem{W01}
W. K. Wootters, 
\textit{Entanglement of formation and concurrence},
Quantum Inf. Comput., \textbf{2} 44 (2001).

\bibitem{MKB04}
F. Mintert, M. Kus, A. Buchleitnar, 
\textit{Concurrence of Mixed Bipartite Quantum States in Arbitrary Dimensions},
Phys. Rev. Lett. \textbf{92}, 167902 (2004).


\bibitem{MSM18}
B. Mukherjee, K. Sengupta, S. N. Majumdar,
\textit{Quantum dynamics with stochastic reset},
Phys. Rev. B \textbf{98}, 104309 (2018).

\bibitem{RTLG18}
D. C. Rose, H. Touchette, I. Lesanovsky, J. P. Garrahan, {\it Spectral properties of simple classical and quantum reset processes},
Phys. Rev. E {\bf 98}, 022129 (2018).

\bibitem{EM11} M. R. Evans, S. N. Majumdar,
\textit{Diffusion with Stochastic Resetting},
Phys. Rev. Lett. {\bf 106}, 160601 (2011).

\bibitem{EM12} M. R. Evans, S. N. Majumdar,
\textit{Diffusion with Optimal Resetting},
J. Phys. A: Math. Theor. {\bf 44}, 435001 (2011).

\bibitem{EMS20} M. R. Evans, S. N. Majumdar, G. Schehr, \textit{Stochastic resetting and
applications}, J. Phys. A: Math. Theor. {\bf 53}, 193001 (2020).


\bibitem{PKR22} A. Pal, S. Kostinski, S. Reuveni, {\it The inspection paradox in stochastic
resetting}, J. Phys. A: Math. Theor. {\bf 55}, 021001 (2022).

\bibitem{NG23}
A. Nagar, S. Gupta,
\textit{Stochastic resetting in interacting particle systems: A review},
J. Phys. A: Math. Theor. \textbf{56}, 283001 (2023).


\bibitem{MV13}
M. Montero, J. Villarroel,
\textit{Monotonous continuous-time random walks with drift and stochastic reset events}, 
Phys. Rev. E \textbf{87}, 012116 (2013).


\bibitem{EM_14}
M. R. Evans, S. N. Majumdar, {\it Diffusion with resetting in arbitrary spatial dimension}, 
J. Phys. A: Math. Theor. {\bf 47}, 285001 (2014).

\bibitem{GMS_14}
S. Gupta, S.~N. Majumdar, G. Schehr, {\it Fluctuating interfaces subject to stochastic resetting},
Phys. Rev. Lett. {\bf 112}, 220601 (2014).

\bibitem{P15}
A. Pal, \textit{Diffusion in a potential landscape with stochastic resetting}, Phys. Rev. E \textbf{91} 012113 (2015).

\bibitem{MSS_15}
S. N. Majumdar, S. Sabhapandit, G. Schehr, {\it Dynamical transition in the temporal relaxation of stochastic processes under resetting}, Phys. Rev. E {\bf 91}, 052131 (2015).

\bibitem{CS_15} C. Christou, A. Schadschneider, {\it Diffusion with resetting in bounded domains}, 
J. Phys. A: Math. Theor. {\bf 48}, 285003 (2015).

\bibitem{MV_16}
M. Montero, J. Villarroel, {\it Directed random walk with random restarts: The Sisyphus random walk}, Phys. Rev. E {\bf 94}, 032132 (2016).

\bibitem{CM16}
V. Mendez, D. Campos,
\textit{Characterization of stationary states in random walks with stochastic resetting},
Phys. Rev. E \textbf{93}, 022106 (2016).

\bibitem{EM16}
S. Eule, J. J. Metzger, \textit{Non-equilibrium steady states of stochastic processes with intermittent resetting},
New J. Phys. \textbf{18}, 033006 (2016).

\bibitem{EM18}
M. R. Evans, S. Majumdar, \textit{Run and tumble particle under resetting: a renewal approach},
J. Phys. A: Math. Theor. \textbf{51} 475003 (2018).

\bibitem{MB18}
G. Mercado-Vasquez, D. Boyer D, \textit{Lotka-Volterra systems with stochastic resetting}, J. Phys. A: Math. Theor. \textit{51} 405601 (2019).

\bibitem{MM19}
J. Masoliver,  M. Montero,
\textit{Anomalous diffusion under stochastic resetting: a general approach},
 Phys. Rev. E \textbf{100} 042103 (2019).

\bibitem{BCS19}
A. S. Bodrova, A. V. Chechkin, I. M. Sokolov, 
\textit{Scaled Brownian motion with renewal resetting}, 
Phys. Rev. E \textbf{100}, 012120 (2019).

\bibitem{MMS_20} M. Magoni, S.~N. Majumdar, G. Schehr, 
{\it Ising model with stochastic resetting},
Phys. Rev. Research {\bf 2}, 033182 (2020).

\bibitem{BLMS_23} M. Biroli, H. Larralde, S. N. Majumdar, and G. Schehr, {\it Extreme Statistics and Spacing 
Distribution in a Brownian Gas Correlated by Resetting}, 
Phys. Rev. Lett. \textbf{130}, 207101 (2023).

\bibitem{TPSRR_20}
O. Tal-Friedman, A. Pal, A. Sekhon, S. Reuveni, Y. Roichman, {\it Experimental Realization of Diffusion with Stochastic Resetting}, 
J. Phys. Chem. Lett. {\bf 11}, 7350 (2020).

\bibitem{BBPMC_20}
B. Besga, A. Bovon, A. Petrosyan, S. N. Majumdar, S. Ciliberto, {\it Optimal mean first-passage time for a Brownian searcher subjected to resetting: Experimental and theoretical results},
Phys. Rev. Research {\bf 2}, 032029(R) (2020).

\bibitem{FBPCM_21}
F. Faisant, B. Besga, A. Petrosyan, S. Ciliberto, S. N. Majumdar, {\it Optimal mean first-passage time of a Brownian searcher with resetting in one and two dimensions: experiments, theory and numerical tests},
J. Stat. Mech. 113203 (2021).


\bibitem{PCML21}
G. Perfetto, F. Carollo, M. Magoni, and I. Lesanovsky,
\textit{Designing nonequilibrium states of quantum matter through stochastic resetting},
Phys. Rev. B \textbf{104}, L180302 (2021).

\bibitem{DDG21} S. Dattagupta, D. Das, S. Gupta,
\textit{Stochastic resets in the context of a tight-binding chain driven by an oscillating field},
J. Stat. Mech. 103210 (2021).

\bibitem{PCL22}
G Perfetto, F. Carollo, I. Lesanovsky,
\textit{Thermodynamics of quantum-jump trajectories of open quantum systems
subject to stochastic resetting}, 
SciPost Phys. {\bf 13}, 079 (2022).

\bibitem{MCPL22}
M. Magoni, F. Carrolo, G. Perfetto, I. Lesanovsky, 
\textit{Emergent quantum correlations and collective behavior in noninteracting
quantum systems subject to stochastic resetting},
Phys. Rev. A {\bf 106}, 052210 (2022).

\bibitem{DCD23}
V. Dubey, R. Chetrite, A. Dhar,
\textit{Quantum resetting in continuous measurement induced dynamics of a qubit},
J. Phys. A: Math. Theor. {\bf 56}, 154001 (2023).

\bibitem{SV23}
F. J Sevilla, A. Vald\'es-Hern\'andez,
\textit{Dynamics of closed quantum systems under stochastic resetting},
J. Phys. A: Math. Theor. {\bf 56} 034001 (2023).

\bibitem{YB23.1}
R. Yin, E. Barkai,
\textit{Restart expedites quantum walk hitting times},
Phys. Rev. Lett. {\bf 130}, 050802 (2023).

\bibitem{YB23.2}
R. Yin, E. Barkai, \textit{Instability in the quantum restart problem},
arXiv preprint: 2301.06100 (2023).

\bibitem{TDFS22}
X. Turkeshi, M. Dalmonte, R. Fazio, and M. Schiro,
\textit{Entanglement transitions
from stochastic resetting of non-Hermitian quasiparticles},
Phys. Rev. B \textbf{105}, L241114 (2022).

\bibitem{S_Book_2011} 
S. Sachdev, 
Quantum phase transitions (Cambridge University Press, 2011).

\bibitem{P70}
P. Pfeuty, 
\textit{The one-dimensional Ising model with a transverse field},
Ann. Phys. \textbf{57}, 79 (1970).

\bibitem{S73}
R. B. Stinchcombe,
\textit{Ising model in a transverse field. I. Basic theory},
J. Phys. C: Solid State Phys. \textbf{6} 2459 (1973).

\bibitem{SRULSS23}
L. Squillante, L. S. Ricco, A. Magnus U., R. E. Lagos-Monaco, A. C. Seridonio, M. de Souza,
\textit{Gruneisen parameter as an entanglement compass},
arXiv:2306.00566


\bibitem{KB87}
K Binder,
\textit{Theory of first-order phase transitions},
Rep. Prog. Phys. \textbf{50} 783 (1987).


\end{thebibliography}
\end{document}